\documentclass[12pt]{article}
\usepackage{amssymb,amsmath,bm,bbm}
\usepackage{epsf}
\usepackage{epsfig}
\usepackage{afterpage}
\usepackage{longtable}
\usepackage{cite}
\usepackage{latexsym, mathrsfs}
\usepackage{axodraw}
\usepackage{graphics}
\usepackage{color}

\DeclareMathOperator{\RE}{Re}
\DeclareMathOperator{\IM}{Im}
\DeclareMathOperator{\diag}{diag}

\def\slash#1{\setbox0=\hbox{$#1$}\dimen0=\wd0
      \setbox1=\hbox{/} \dimen1=\wd1 \ifdim\dimen0>\dimen1
      \rlap{\hbox to \dimen0{\hfil/\hfil}} #1                        \else
      \rlap{\hbox to \dimen1{\hfil$#1$\hfil}}
      /   \fi}

\newcommand{\lsim}{
\mathrel{\hbox{\rlap{\hbox{\lower4pt\hbox{$\sim$}}}\hbox{$<$}}}}

\newcommand{\gsim}{
\mathrel{\hbox{\rlap{\hbox{\lower4pt\hbox{$\sim$}}}\hbox{$>$}}}}

\def\eps{\varepsilon}

\newcommand{\tev}{\, {\rm TeV}}
\newcommand{\gev}{\, {\rm GeV}}
\newcommand{\mev}{\, {\rm MeV}}

\newcommand{\Heff}{{\cal H}_\text{eff}}

\def\beq{\begin{equation}}
\def\eeq{\end{equation}}

\newcommand{\be}{\begin{equation}}
\newcommand{\ee}{\end{equation}}
\newcommand{\bea}{\begin{eqnarray}}
\newcommand{\eea}{\end{eqnarray}}
\newcommand{\nn}{\nonumber}
\newcommand{\bi}{\begin{itemize}}
\newcommand{\ei}{\end{itemize}}
\newcommand{\ord}{{\cal O}}

\newcommand{\newsection}[1]{\section{#1}\setcounter{equation}{0}}

\definecolor{lightgray}{gray}{.5}

\begin{document}
\begin{titlepage}
\vspace*{-0.5truecm}

\begin{flushright}
{TUM-HEP-698/08}\\
MPP-2008-115
\end{flushright}

\vfill

\begin{center}
\boldmath

{\Large\textbf{$\Delta F=2$ Observables and Fine-Tuning in a \vspace{.2cm}\\Warped Extra Dimension with Custodial Protection}}
\unboldmath
\end{center}

\vspace{0.4truecm}

\begin{center}
{\bf  Monika Blanke$^{a,b}$, Andrzej J.~Buras$^{a,c}$, Bj\"orn Duling$^a$,\\  Stefania Gori$^{a,b}$ and Andreas Weiler$^{d}$}
\vspace{0.4truecm}

{\footnotesize
 $^a${\sl Physik Department, Technische Universit\"at M\"unchen,
D-85748 Garching, Germany}\vspace{0.2cm}

 {\sl $^b$Max-Planck-Institut f{\"u}r Physik (Werner-Heisenberg-Institut), \\
D-80805 M{\"u}nchen, Germany}\vspace{0.2cm}

 $^c${\sl TUM Institute for Advanced Study, Technische Universit\"at M\"unchen,  Arcisstr.~21,\\ D-80333 M\"unchen, Germany}\vspace{0.2cm}

 $^{d}${\sl Institute for High Energy Phenomenology,\\
Newman Laboratory of Elementary Particle Physics,\vspace{-0.1cm}\\
Cornell University, Ithaca, NY 14853, USA } }

\end{center}
\begin{abstract}
\noindent
We present a complete study of $\Delta S=2$ and $\Delta B=2$ processes 
in  a warped extra dimensional model with a custodial protection of $Zb_L\bar b_L$,
 including $\varepsilon_K$, $\Delta M_K$,  $\Delta M_s$, $\Delta M_d$, $A_{\rm
   SL}^q$, $\Delta\Gamma_q$, $A_{\rm CP}(B_d \rightarrow \psi K_S)$ and
 $A_{\rm CP}(B_s \rightarrow \psi \phi)$. These processes are affected by tree
 level contributions from Kaluza-Klein gluons, the heavy KK photon,
   new heavy electroweak gauge bosons $Z_H$ and $Z'$, {and in principle by
     tree level $Z$ contributions.}
We confirm recent findings that the fully anarchic approach where all the hierarchies in quark masses and
weak mixing angles are geometrically explained seems implausible and we confirm that the KK mass scale $M_\text{KK}$ generically has to be at least {$\sim 20\tev$} to satisfy the $\eps_K$ constraint. We point out, however, that there exist regions in parameter space with only modest fine-tuning in the 5D Yukawa couplings which satisfy all existing $\Delta F=2$ and electroweak precision constraints for scales $M_\text{KK}\simeq 3\tev$ in reach of the LHC.  Simultaneously we find that $A_\text{CP}(B_s\to\psi\phi)$ and $A^s_\text{SL}$ can be much larger than in the SM as indicated by recent results from CDF and D{\O} data. 
We point out that for $B_{d,s}$ physics {$\Delta F=2$ observables the
  complex $(Z_H,Z')$} can 
compete with KK gluons, while the {tree level $Z$ and KK photon contributions are
very small. In particular we point out that the $Zd^i_L\bar d^j_L$ couplings
are protected by the
 custodial symmetry.}
As a by-product we show the relation of the RS flavour model to the Froggatt-Nielsen mechanism and we provide analytic formulae for the effective flavour mixing matrices in terms of the fundamental 5D parameters.

\end{abstract}

%
%
%
\end{titlepage}

\setcounter{page}{1}
\pagenumbering{arabic}

\newsection{Introduction}\label{sec:int}

The Standard Model of particle physics is in spectacular agreement
with everything we know about interactions of elementary particles. Yet
it requires large hierarchies to be put in by hand. There is no explanation for the hierarchy between the electroweak (EW) scale and the Planck scale and for the observed hierarchical pattern of fermion masses and mixings.

%
%

{Among the} most ambitious proposals {to explain these hierarchies} are models with a warped extra spatial dimension first proposed by Randall and Sundrum (RS) \cite{Randall:1999ee} where the SM fields, except the Higgs boson, are allowed to propagate in the bulk \cite{Gherghetta:2000qt,Chang:1999nh,Grossman:1999ra}. These models provide a geometrical explanation of the hierarchy between the Planck scale and the  EW scale and one can naturally generate the  hierarchies in the fermion mass spectrum and mixing angles~\cite{Grossman:1999ra,Gherghetta:2000qt} while simultaneously suppressing flavour changing neutral current (FCNC) interactions~\cite{Huber:2003tu,Agashe:2004cp}. Recently realistic models of EW symmetry breaking (EWSB) have been constructed~\cite{Agashe:2003zs,Csaki:2003zu,Agashe:2004rs,Cacciapaglia:2006gp,Contino:2006qr,Carena:2006bn} and one can even achieve gauge coupling unification~\cite{Agashe:2002pr,Agashe:2005vg}. 

In this work we discuss the flavour structure of models based on the bulk gauge group 
{\be\label{eq:1.1}
G_\text{bulk}=SU(3)_c\times SU(2)_L\times SU(2)_R\times U(1)_X\times P_{LR}\,.
\ee}
The SM fermions are embedded in representations of $G_\text{bulk}$, so that there is a protection of the $T$ parameter {\cite{Agashe:2003zs,Csaki:2003zu}} and the coupling $Z b_L \bar{b}_L$ {\cite{Agashe:2006at}}. This allows KK masses of order $M_\text{KK} \simeq (2-3)\tev$ which are in the reach of the LHC {\cite{Contino:2006qr,Cacciapaglia:2006gp,Carena:2007ua,Djouadi:2006rk,Bouchart:2008vp}}.

The goal of the present paper is to analyse the well measured FCNC processes related to particle-antiparticle mixings $K^0-\bar K^0$ and $B_{d,s}^0-\bar B_{d,s}^0$. The off-diagonal mixing amplitudes $M_{12}^i$ ($i=K, d, s$) receive dangerous tree level contributions from Kaluza-Klein (KK) gluon and {EW gauge boson} exchanges \cite{Burdman:2003nt,Agashe:2004cp}. We would like to know whether this model can be made consistent with simultaneous constraints {from  $\varepsilon_K$}, $\Delta M_K$,  $\Delta M_d$, $\Delta M_s$ and the mixing induced CP-asymmetry $S_{\psi K_S}$ for KK scales as low as $M_\text{KK} \simeq (2-3)\tev$.

A recent study \cite{Csaki:2008zd}, which applied model-independent results of the UTfit group~\cite{Bona:2007vi} to RS-type models, concluded that the measured value of $\varepsilon_K$ implies that the lightest KK gluon mode has to be generically heavier than $\sim 21\tev$, if the hierarchy of fermion masses and weak mixings is solely due to geometry and the 5D Yukawa couplings are anarchic and of $\mathcal{O}(1)$.
KK particles {that heavy} undermine the basic motivation for RS models. We would like to investigate if the KK scale could be lowered to be in reach of the LHC by allowing for a modest hierarchy and some tuning in the fundamental 5D Yukawa couplings.

The dominant flavour constraint comes from the CP-violating contribution to chirality flip operators $\mathcal{Q}_{LR}$ which are very strongly suppressed in the SM, but present in RS models. The lower bound on the KK gluon mass obtained in~\cite{Csaki:2008zd} originates from the excessive contribution of $\mathcal{Q}_{LR}$ to $\eps_K$. One of our strategies will be to find regions of parameter space consistent with EW {precision observables} {\cite{Contino:2006qr,Cacciapaglia:2006gp,Carena:2007ua,Djouadi:2006rk,Bouchart:2008vp}} for which   $\mathcal{Q}_{LR}$ is sufficiently suppressed even if the 5D Yukawa couplings
are mostly anarchic.

Several alternative models have been proposed to deal with the flavour problem of RS. All depart
from the fully anarchic set-up by incorporating
some sort of flavour symmetry. One 
approach is to protect the model from all tree
level FCNCs by incorporating a full 5D GIM mechanism~\cite{Cacciapaglia:2007fw}. The bulk respects a $U(3)^3$ flavour symmetry and all flavour mixing is generated
by kinetic terms on the UV brane. Although this model 
is safe, since its effective theory is minimal flavour violating (MFV) \cite{D'Ambrosio:2002ex,Chivukula:1987py,Hall:1990ac,Buras:2000dm,Buras:2003jf}, it leaves the origin 
of the large hierarchies in the flavour sector unanswered.
More recent proposals therefore seek to suppress dangerous FCNCs and  simultaneously try to explain the hierarchical structure of the flavour sector.
One of them is the so called "5D MFV" model~\cite{Fitzpatrick:2007sa}. Here one postulates that the only sources of flavour breaking are two anarchic Yukawa spurions. The low-energy limit is not MFV, and
the additional assumption,
that brane and bulk terms in the down sector are effectively aligned, is needed to suppress dangerous FCNCs. Recently, an economical model has been proposed~\cite{Santiago:2008vq} in which one assumes a $U(3)$ flavour symmetry for the 5D fields containing the right handed down quarks. This global symmetry forces the couplings of the right handed down quarks to the vector KK modes to be degenerate.	Dangerous contributions to $\mathcal{Q}_{LR}$ are only generated by suppressed mass insertions on the IR brane where the symmetry is necessarily  broken (see \cite{Csaki:2008eh} for a discussion of possibly problematic fermionic brane kinetic terms).
Another recent approach \cite{Csaki:2008eh} presents a simple model where the key ingredient are two horizontal $U(1)$ symmetries. The SM fields are embedded  into the 5D fields motivated by protecting $Zb_L\bar b_L$. The horizontal $U(1)$ symmetries force an alignment of bulk masses and down Yukawas which strongly suppresses FCNCs in the down sector. FCNCs in the up sector, however, can be close to experimental limits. In {the present paper, however,} we will study the original version of the model.

As there have been other analyses of particle-antiparticle mixing in the RS model in the past {\cite{Burdman:2002gr,Burdman:2003nt,Agashe:2004ay,Agashe:2004cp,Moreau:2006np,Chang:2006si}}, most recently in \cite{Csaki:2008zd}, it is mandatory for us to state what is new in our paper:
\bi
\item
First of all we perform a simultaneous analysis of the most interesting $\Delta F=2$ observables in the $K$ and $B_{d,s}$ meson systems in conjunction with $\eps_K$. In \cite{Csaki:2008zd} only one Wilson coefficient at a time has been considered. This will give us a global picture of correlations between various observables. Such an analysis has not been performed in the literature so far.
\item
Similarly we perform the full renormalisation group analysis at the NLO level, including not only the two $\mathcal{Q}_{LR}$ operators in our analysis, but also $\mathcal{Q}_{LL}$ and $\mathcal{Q}_{RR}$. We would like to emphasise that this is essential since the operator $\mathcal{Q}_{LL}$, although subleading in $\eps_K$, turns out to be as important as $\mathcal{Q}_{LR}$ in $B_{d,s}$ physics observables. On the other hand $\mathcal{Q}_{RR}$ is subdominant in all processes considered in this paper.
\item
In addition to tree level KK gluon exchanges considered in \cite{Csaki:2008zd} we present {for the first time} the formulae for the EW tree level contributions of $Z, Z', Z_H$ and the KK photon $A^{(1)}$ to the {$\Delta F=2$} Wilson coefficients of the operators involved. {Quite unexpectedly we find that the $Z'$ and $Z_H$ contributions, while subleading with respect to KK gluon contributions in the case of $\varepsilon_K$ and $\Delta M_{K}$, can compete with the latter in the case of $B_{s,d}$ physics observables. The contributions of the KK photon turn out to be small.
\item
We point out and demonstrate explicitly that in the model in question {tree level {flavour violating} $Z$ couplings to left-handed down-type quarks are strongly
suppressed by the $P_{LR}$ symmetry up to small symmetry breaking effects due to the UV boundary conditions.}} {This suppression mechanism works not only for the KK gauge boson contribution, but also for the KK fermion contribution to the $Z$ coupling, as the fermion representations are symmetric under $P_{LR}$.}
\item
We show that it is possible to simultaneously fit the SM quark masses and CKM parameters within their experimental $2\sigma$ ranges and obtain agreement with all available constraints on $\Delta F=2$ observables.
\item
We present a new useful parameterisation of the 5D Yukawa coupling matrices, taking into account only physical parameters.
\item
As a by-product we analyse the connection of RS models to the Froggatt-Nielsen mechanism \cite{Froggatt:1978nt} and provide analytic formulae for the effective flavour mixing matrices in terms of the fundamental 5D parameters.
\ei

Our paper is organised as follows. In Section \ref{sec:model} we summarise  briefly the main ingredients of the {model} in question. Readers familiar with RS models may skip this section and start directly with Section \ref{sec:FN}, where we analyse the connection of flavour in RS models with the Froggatt-Nielsen mechanism  and derive explicit formulae for the effective flavour mixing matrices $\mathcal{U}_{L,R}$ and $\mathcal{D}_{L,R}$ in terms of the fundamental 5D parameters. Then in Section \ref{sec:trans} we derive the effective Hamiltonians for $K^0-\bar K^0$, $B_{d}^0-\bar B_{d}^0$ and $B_{s}^0-\bar B_{s}^0$ mixings originating from tree-level KK gluon exchange and we calculate most interesting observables such as the CP-violating parameter $\varepsilon_K$, the mass differences $\Delta M_K$,  $\Delta M_d$ and $\Delta M_s$, the CP-asymmetries $A_{\rm SL}^q$ ($q=d,s$), $A_{\rm CP}(B_d \rightarrow \psi K_S)$ and $A_{\rm CP}(B_s \rightarrow \psi \phi)$ and the width difference $\Delta\Gamma_q$. {We also give formulae for EW tree level contributions and estimate their size. Interestingly in the case of $B_{d,s}$ physics observables some of these contributions can compete with the KK gluon contributions.} {We also demonstrate that tree level flavour violating $Z$ couplings to left-handed down-type quarks are strongly suppressed in the model in question, due to the custodial protection mechanism. This finding has important implications not only for $\Delta F=2$ processes, but also for $\Delta F=1$ rare decays.} In Section \ref{sec:strategy} we outline our strategy for the numerical analysis, presenting in particular a useful parameterisation for the 5D Yukawa couplings. In Section \ref{sec:num} a detailed numerical analysis of  particle-antiparticle mixing observables is presented.  We summarise our results in Section \ref{sec:conc}. 

\newsection{The Model}\label{sec:model}

\subsection{Geometric Setup}

The class of models we are considering is based on the Randall-Sundrum (RS) geometric background, i.\,e. we consider a 5D spacetime, where the extra dimension is compactified to the interval $0\le y\le L$, with a warped metric given by
\cite{Randall:1999ee}
\be\label{eq:RS}
ds^2=e^{-2ky}\eta_{\mu\nu}dx^\mu dx^\nu - dy^2\,.
\ee
Here the curvature scale $k$ is assumed to be $k\sim\ord(M_\text{Pl})$. Due to the exponential warp factor $e^{-ky}$, the effective energy scales depend on the position $y$ along the extra dimension.
In order to obtain a solution to the gauge hierarchy problem, we set $e^{kL}=10^{16}$ and treat
\be\label{eq:f}
f=ke^{-kL}\sim \ord(\text{TeV})
\ee
as the only free parameter coming from space-time geometry.

\subsection{KK Gluons}

The main actors in $\Delta F=2$ processes in RS are {at first sight} KK gluons originating from the bulk $SU(3)_c$ symmetry, and in particular the first KK excitation. Therefore let us restrict ourselves to this mode, to be simply called KK gluon, in what follows. The profile of the KK gluon along the extra dimension is given by \cite{Gherghetta:2000qt}
\be\label{eq:gaugeprofile}
g(y)= \frac{e^{ky}}{N}\left[J_1\left(\frac{M_\text{KK}}{k}e^{ky}\right)+b Y_1\left(\frac{M_\text{KK}}{k}e^{ky}\right)\right]\simeq \frac{e^{ky}}{N}J_1\left(\frac{M_\text{KK}}{k}e^{ky}\right)\,,
\ee
where $J_1(x)$ and  $Y_1(x)$ are the Bessel functions of first and second kind, $b\simeq 0$ is determined by the boundary conditions at $y=0,L$ and $N$ is a normalisation constant.
The KK mass $M_\text{KK}$ can be numerically determined to be \cite{Agashe:2007ki}
\be
M_\text{KK} \simeq 2.45  f\,,
\ee
with $f$ being defined in \eqref{eq:f}. {We would like to caution the reader that a different notation has been used in \cite{Casagrande:2008hr}: Their $M_\text{KK}$ corresponds to our $f$, so that  in spite of comparable $M_\text{KK}$  the masses of
the first KK gauge bosons in that paper are  larger  than in our analysis.}

It is crucial to note that the KK gluon bulk profile is not flat along the extra dimension, but due to the factor $e^{ky}$ strongly localised towards the IR brane. This localisation will give rise to flavour non-universal couplings of the KK gluon and eventually to tree level FCNC transitions, as we will discuss later on.

{
\subsection{Electroweak Gauge Sector}

The other actors in our {analysis} are the neutral EW gauge bosons $Z,Z_H,Z'$ and the KK photon $A^{(1)}$ \cite{Blanke:2008aa}, originating from the enlarged gauge group
\be
SU(2)_L\times SU(2)_R\times U(1)_X\times P_{LR}\,.
\ee
We give here for the first time their contributions to $\Delta F=2$ processes using the {couplings worked} out in \cite{Blanke:2008aa}. While subleading in the case of $\eps_K$ and $\Delta M_K$ they turn out to play {an important} role in the case of $B_{d,s}$ physics observables.

}

\subsection{Bulk Fermionic Zero Modes}

Bulk fermions in RS models
offer a natural explanation of the observed hierarchies in fermion masses and mixings \cite{Grossman:1999ra,Gherghetta:2000qt,Huber:2003tu} and provide at the same time a powerful suppression mechanism for FCNC interactions, the so-called \emph{RS-GIM mechanism} \cite{Agashe:2004cp}.

The bulk profile of a fermionic zero mode depends strongly on its bulk mass parameter $c_\Psi$. In case of a left-handed zero mode $\Psi_L^{(0)}$ it is given by {\cite{Grossman:1999ra,Gherghetta:2000qt}}
\be
f^{(0)}_L(y,c_\Psi) = \sqrt{\frac{(1-2c_\Psi)kL}{e^{(1-2c_\Psi)kL}-1}} e^{- c_\Psi ky}\label{eq:fL}
\ee
with respect to the warped metric.
Therefore, for $c_\Psi >1/2$ the fermion $\Psi_L^{(0)}$ is localised towards the UV brane and exponentially suppressed on the IR brane, while for $c_\Psi < 1/2$ it is localised towards the IR brane. The bulk profile for a right-handed zero mode $\Psi_R^{(0)}$ can be obtained from
\be
f^{(0)}_R(y,c_\Psi) = f^{(0)}_L(y,-c_\Psi)\,,\label{eq:fR}
\ee
so that its localisation depends on whether $c_\Psi <-1/2$ or $c_\Psi >-1/2$. Note that the left- and right-handed zero modes present in the spectrum necessarily belong to different 5D multiplets, so that generally $c_{\Psi_L} \ne c_{\Psi_R}$.

In order to reproduce the SM quark content in the low energy limit, three left-handed zero mode {$SU(2)_L$} doublets $Q^i_L$ with bulk mass parameters $c_Q^i$ and three right-handed zero mode up- and down-type {$SU(2)_L$} singlets $u^j_R$ and $d^j_R$ with bulk mass parameters $c_{u,d}^j$, respectively, are required.
As the KK fermions do not enter directly tree level $\Delta F=2$ processes, we do not specify the fermion representations here. They are discussed in detail in \cite{Blanke:2008aa}. {We stress however that in order to preserve the custodial $Z d_L^i \bar d_L^j$ protection, $P_{LR}$ symmetric fermion representations are required.}

The coupling of a zero mode fermion $\Psi^{(0)}_{L,R}$ to the KK gluon $G^{(1)a}_\mu$ in the flavour eigenbasis is given by
\be\label{eq:vertex}
\bar \Psi^{(0)}_{L,R} G^{(1)a}_\mu \Psi^{(0)}_{L,R} \quad : \qquad -i\gamma^\mu t^a \frac{g_s^\text{5D}}{L^{3/2}}\int_0^L dy\,e^{ky} \left[f^{(0)}_{L,R}(y,c_\Psi)\right]^2 g(y)\,,
\ee
with $g_s^\text{5D}$ being the 5D $SU(3)_c$ gauge coupling constant and $t^a$ the $SU(3)_c$ generators. The 4D QCD coupling constant $g_s^\text{4D}$ is then given by
\be\label{eq:match}
g_s^\text{4D} =\frac{1}{p_\text{UV}}\frac{g_s^\text{5D}}{\sqrt{L}}\,,
\ee
where in the absence of brane kinetic terms $p_\text{UV}\equiv 1$.

 Note that flavour universality is strongly violated due to the dependence of the overlap integral on the bulk mass parameter $c_\Psi$.

\subsection{Brane Kinetic Terms}\label{sec:BKT}

One should keep in mind that localised brane kinetic terms for the gluon can change the relation
between the bulk gauge coupling $g_s^\text{5D}$ and the 4D QCD gauge coupling.  The matching relation at a given scale contains both a bulk term and contributions from IR and UV brane kinetic terms. The UV brane terms consist of a possible (positive) bare kinetic term and an always present negative 
term~\cite{Randall:2001gb,
Goldberger:2002cz,Goldberger:2002hb,Goldberger:2003mi,Choi:2002ps,Agashe:2002bx,Agashe:2002pr,Contino:2002kc}, which is due to the asymptotically free QCD running from the Planck scale to the TeV scale, see~\cite{Csaki:2008zd} for an extended discussion. 
The running of the IR brane kinetic term is negligible and we will therefore focus on the UV brane localised kinetic terms.
One possibility is that there are no bare UV brane kinetic terms at the Planck scale and that the negative contribution from the running reduces the bulk gauge coupling to {$g_s^\text{5D}\sqrt{k} \approx 3$}, corresponding to $p_\text{UV}\approx 1/2$. Another possibility is that large brane kinetic terms at the
Planck scale are present which would render the bulk strongly coupled, {$g_s^\text{5D}\sqrt{k} \sim 4 \pi$}. 
Usually one assumes an intermediate scenario where the bare kinetic 
terms are of  exactly the same size as the one induced by the one-loop running.
This cancellation is assumed in the trivial matching relation~(\ref{eq:match}) for $p_\text{UV}\equiv 1$ and results in {$g_s^\text{5D}\sqrt{k} \approx 6$} for the QCD bulk gauge coupling.  In our analysis we have set $p_\text{UV}\equiv 1$ which in~\cite{Csaki:2008zd} resulted in $M_\text{KK}>21\tev$ in order to be consistent with the experimental value of $\eps_K$.

{Similarly, also in the electroweak sector, brane kinetic terms can be present and thus alter the simple tree level matching condition $g^\text{4D} = {g^\text{5D}}/{\sqrt{L}}$.
Here the situation is additionally complicated by the fact that on the UV brane the gauge group is broken to $SU(2)_L\times U(1)_Y$. Therefore different UV brane kinetic terms can be present for $SU(2)_L$ and $U(1)_Y$, so that in general $p_\text{UV}^{SU(2)_L} \ne p_\text{UV}^{U(1)_Y} \ne 1$ and also different from $p_\text{UV}$ in the strong sector. 
Consequently the relative size of strong and electroweak contributions to $\Delta F=2 $ processes depends on the size of possible brane kinetic terms.
In order to allow for an easy comparison of the KK electroweak gauge boson effects with the KK gluon effects, we will, as in the former case, assume also here the intermediate scenario $p_\text{UV}^{SU(2)_L} \approx p_\text{UV}^{U(1)_Y} \approx 1$. In order to keep the analytic expressions in Section \ref{sec:EWcont} simple, we omit $p_\text{UV}^{SU(2)_L}, p_\text{UV}^{U(1)_Y}$ in the formulae. A generalisation to include these terms is straightforward.

}

\subsection{Higgs Field and Yukawa Couplings}\label{sec:2.4}

The present analysis does not require the specification of the exact EWSB mechanism. Instead we will simply assume the presence of a Higgs doublet $H(x^\mu)$ residing  on  the IR brane. Once its neutral component develops a VEV $v\simeq 246\gev$, 
EWSB is achieved.

The effective 4D Yukawa couplings, relevant for the SM fermion masses and mixings, are then given by 
\be\label{eq:Yud}
Y^{u,d}_{ij} = \lambda^{u,d}_{ij}\,\frac{e^{kL}}{kL} f^{(0)}_{L}(y=L,c_Q^i)  f^{(0)}_{R}(y=L,c_{u,d}^j) \equiv \lambda^{u,d}_{ij}\,\frac{e^{kL}}{kL} f^Q_i  f^{u,d}_j \,,
\ee
where $\lambda^{u,d}$ are the fundamental 5D Yukawa coupling matrices. In order to preserve perturbativity of the model, we require, as usual, $|\lambda^{u,d}_{ij}|\le3$. Here and in the following, we work in the special basis in which the bulk mass matrices are taken to be real and diagonal. Such a basis can always be reached by appropriate unitary transformations in the $Q_i$, $u_i$ and $d_i$ sectors.

Due to the exponential dependence of $Y^{u,d}$ on the bulk mass parameters $c_{Q,u,d}$, the strong hierarchies of quark masses and mixings can be traced back to $\ord(1)$ bulk masses and anarchic 5D Yukawa couplings  $\lambda^{u,d}$. We will elaborate more on this issue in the next section.

The transformation from the quark flavour eigenbasis $\tilde u_{L,R},\tilde d_{L,R}$ to the mass eigenbasis $ u_{L,R}, d_{L,R}$ will then, as in the SM, be performed by means of four unitary mixing matrices ${\cal U}_{L,R},{\cal D}_{L,R}$, where 
\begin{gather}
u_L = \mathcal{U}_L^\dagger\tilde u_L\,,\qquad 
u_R = \mathcal{U}_R^\dagger\tilde u_R\,, \label{eq:ULR}\\
d_L = \mathcal{D}_L^\dagger\tilde d_L\,,\qquad 
d_R = \mathcal{D}_R^\dagger\tilde d_R\,, \label{eq:DLR}
\end{gather}
and the CKM matrix is given by
\be
V_\text{CKM}= \mathcal{U}_L^\dagger\mathcal{D}_L\,.
\ee

{
As argued in \cite{Agashe:2004cp,Casagrande:2008hr}, the mixing of fermion zero modes with their heavy KK resonances induces flavour violating couplings of the Higgs boson\footnote{We would like to thank Uli Haisch for bringing this issue to our attention.}, eventually leading to additional tree level contributions to $\Delta F=2$ processes. However it can straightforwardly be seen (see Appendix~\ref{app:Higgs} for details) that these contributions are strongly suppressed in the model in question. Therefore we will not consider them any further.
}
{

\subsection{Impact of Higher KK Fermion Modes}\label{sec:KKfermions}

Even at tree level higher KK fermion modes affect flavour observables through mixing with SM fermions. Depending on the particular structure of the Yukawa interactions, like-charged fermions of any KK level mix with each other. The relevant three-by-three subsets of the infinite dimensional full rotation matrices necessarily deviate from unitarity. Besides that, the small admixture of higher KK fermion modes to SM fermions modifies their {gauge couplings} since SM fermions and KK fermion modes couple {in general differently to the various gauge boson modes. This is true not only for the heavy KK gauge bosons, but in particular also for the $Z$ boson, as fermions with different weak isospin mix with each other.} 

{As the KK fermion mixing appears as a consequence of EWSB, the corrections to the zero mode gauge couplings can be estimated to be of order $\ord(v^2/M_\text{KK}^2)$. While this is a sub-leading and therefore small effect in the case of KK gauge boson couplings, whose flavour violating couplings are $\ord(1)$, the situation is a priori different in the case of flavour violating $Z$ couplings that appear first at the $\ord(v^2/M_\text{KK}^2)$ level\footnote{We would like to thank Csaba Csaki for bringing this issue to our attention.}. However we have checked numerically by including the first fermionic KK excitations that the KK fermion contribution to the $Z$ coupling is generally suppressed with respect to the $(Z^{(1)}, Z_X^{(1)})$ contribution. Therefore their impact is subleading not only in the case of $\Delta F=2$ observables, where the $Z$ contributions are of higher order anyway, but also in the case of $\Delta F=1$ rare decays studied in \cite{BBDGG}. In this context we underline that the custodial protection mechanism for the $Z d_L^i \bar d_L^j$ coupling discussed in Section \ref{sec:Z} is effective not only for the $(Z^{(1)}, Z_X^{(1)})$ contributions, but also for the KK fermion contributions, as the fermions in the model considered are embedded in $P_{LR}$-symmetric representations.}

The explicit analytic formulae for including these effects to arbitrary order in KK excitations clearly are beyond the scope {of this paper} and will be presented elsewhere. In the course of the present analysis we checked numerically that the effects of the first KK excitations on $\Delta F=2$ observables amount to at most 10\% for a wide majority of generated data points. For completeness sake we mention that for a very small subset of data points, that display a large fine-tuning in some observable, the effect on that particular observable can be $\mathcal{O}(1)$. {This is clearly due to the accidental suppression of the leading contribution which makes the sub-leading 
corrections relatively larger.}
Since the aim of the present work is to look for areas in parameter space with acceptable fine-tuning only, ignoring these points is justifiable in the framework of our analysis.

Having at hand these results, we like to underline that the inclusion of the first fermion KK modes neither does modify the overall picture of $\Delta F=2$ observables in the model under consideration, nor does it affect the results stated in the following sections. In the light of other theoretical uncertainties, e.\,g. from the exchange of higher {KK gauge} excitations, that is also at the 10\% level, or the uncertainties stemming from brane kinetic terms, we believe that our treatment of KK fermion modes is fully sufficient for the time being.

}

\newsection{Geometric Origin of Masses and Mixing Angles}\label{sec:FN}

\subsection{Analogy between RS and Froggatt-Nielsen Scenarios}

The aim to explain the observed hierarchies in the fermion masses and their flavour mixing matrices by making use of approximate flavour symmetries traces back to the late 1970s, to the well-known work of Froggatt and Nielsen \cite{Froggatt:1978nt}. In that pioneering paper a global $U(1)_F$ flavour symmetry has been introduced, under which the various quark fields carry different charges while the SM Higgs $H$ is neutral under $U(1)_F$. In order to allow for non-vanishing flavour mixing, the flavour symmetry is spontaneously broken by the VEV of a scalar $\Phi$, {the so-called flavon field}, that transforms as gauge singlet, but is (singly) charged under $U(1)_F$. {In order to obtain small flavour violating effects consistent with observation, 
the flavon VEV $\langle \Phi \rangle$ has to be much smaller than its mass $m_\Phi \sim\Lambda$. The effective flavour violating parameter is then given by $\epsilon = \langle \Phi \rangle /\Lambda \ll 1$.}

A very similar structure is encountered in RS models with fermions living in the 5D bulk. In that case the flavour $U(1)_F$ symmetry corresponds to translations along the extra-dimensional coordinate $0\le y\le L$, under which the metric is self-similar. The Higgs field, living on the IR brane, is external to this self-similarity of the bulk. The fermions, on the other hand, are localised along the extra dimension by means of their bulk mass parameters $c_{Q,u,d}$, i.\,e.  the bulk mass parameters can be interpreted as charges under self-similarity transformations. Self-similarity is broken explicitly by the presence of the IR brane, giving rise to the symmetry breaking parameter $e^{-kL}\ll 1$. The one-to-one correspondence between a Froggatt-Nielsen flavour symmetry and bulk fermions in RS is summarised in Table~\ref{tab:FN}.

\begin{table}[ht]
\begin{center}
{\small\renewcommand{\arraystretch}{1.2}
\begin{tabular}{|c|c| c|}
\hline
Froggatt-Nielsen symmetry & bulk fermions in RS & CFT interpretation \\ \hline
$U(1)_F$ symmetry & self-similarity along $y$ & scale invariance \\
$U(1)_F$ charges $Q_F=a_i,b_i,d_i$ & bulk mass parameters $c_{Q,u,d}^i$ & \renewcommand{\arraystretch}{.7}
\begin{tabular}{c}
anomalous dimensions $\gamma_i$\\
of fermionic operators
\end{tabular} \renewcommand{\arraystretch}{1}\\
VEV of scalar $\Phi$ ($Q_F=1$) & IR brane at $y=L$ & CFT condensate \\
{$\epsilon = \langle \Phi \rangle / \Lambda \ll 1$} & warp factor $e^{-kL}$ & scale ratio $\Lambda_\text{IR}/M_\text{Pl}$\\
\hline
\end{tabular}\renewcommand{\arraystretch}{1}}
\end{center}
\caption{\label{tab:FN}\it Correspondence between Froggatt-Nielsen flavour symmetry, bulk fermions in RS and their dual CFT description.}
\end{table}

It is also interesting to consider this correspondence  in the CFT holographic dual. In that picture   the 5D bulk coordinate $y$ corresponds to the energy scale of the CFT. Translational invariance along the bulk then translates directly into scale invariance. This scale invariance is spontaneously broken by a CFT condensate at the $\Lambda_\text{IR}\sim1\tev$ scale, corresponding to the IR brane at $y=L$. The symmetry breaking parameter is therefore given by $\Lambda_\text{IR}/M_\text{Pl}\ll 1$, where the large hierarchy between these two scales arises naturally as the spontaneous breaking of the CFT is due to radiative corrections. 

As the Higgs field is part of the conformal sector, it couples strongly to composite fermionic operators of anomalous dimension $\gamma_i$, one for each quark species. Those fermionic operators then mix with elementary fermions that are external to the CFT and correspond to the SM quarks. The size of this mixing is directly related to the effective Yukawa couplings and depends exponentially on the anomalous dimensions  $\gamma_i$. Thus effectively the $\gamma_i$ can be interpreted as different flavour ``charges''.

The structure of the effective Yukawa coupling matrices in RS then turns out to be completely analogous to that analysed by Froggatt and Nielsen \cite{Froggatt:1978nt}\footnote{This has also been noticed and worked out independently in \cite{Casagrande:2008hr}.}. This analogous structure can now be used to derive analytic expressions for the quark masses and flavour mixing matrices $\mathcal{U}_{L,R}, \mathcal{D}_{L,R}$ in terms of the fundamental model parameters. Therefore we have checked the respective analytic expressions in \cite{Froggatt:1978nt} and carefully adapted them to the RS scenario in question. The result is summarised in the next section.

\subsection{Quark Masses and Flavour Mixing Made Explicit}

Assuming an IR brane localised Higgs boson the effective 4D Yukawa couplings $Y^{u,d}$ can be written in terms of the fundamental 5D Yukawa couplings $\lambda^{u,d}$ and the fermion shape functions $f^Q_i,f^u_i,f^d_i$ $(i=1,2,3)$ as given in \eqref{eq:Yud},
where the hierarchies in the 4D Yukawas arises through the hierarchies
\begin{gather}
f^{Q}_1 \ll f^{Q}_2 \ll f^{Q}_3\,,\\
f^{u}_1 \ll f^{u}_2 \ll f^{u}_3\,,\\
f^{d}_1 \ll f^{d}_2 \ll f^{d}_3\,.
\end{gather}

Keeping only the leading terms in the hierarchies  $f^{Q,u,d}_i/f^{Q,u,d}_j$ $(i<j)$, we obtain for the quark masses\footnote{We would like to thank Katrin Gemmler for checking all formulae given in this section and Appendix~\ref{app:A}.}
\bea
m_b &=& \frac{v}{\sqrt{2}}  \lambda^d_{33}
\frac{e^{kL}}{kL} f^Q_3 f^d_3\,,\label{eq:mbFN}\\
m_s &=& \frac{v}{\sqrt{2}}  \frac{\lambda^d_{33}\lambda^d_{22}-\lambda^d_{23}\lambda^d_{32} }{\lambda^d_{33}}
\frac{e^{kL}}{kL} f^Q_2 f^d_2\,,\\
m_d &=& \frac{v}{\sqrt{2}}  \frac{\det(\lambda^d)}{\lambda^d_{33}\lambda^d_{22}-\lambda^d_{23}\lambda^d_{32} } 
 \frac{e^{kL}}{kL}f^Q_1 f^d_1\,,\label{eq:mdFN}
\eea
and analogous expressions for the up-type quark masses $m_{t,c,u}$, with only replacing ``$\lambda^d$'' by ``$\lambda^u$'' and ``$f^d$'' by ``$f^u$''.. 

Similarly, for the flavour mixing matrices ${\cal U}_{L,R},{\cal D}_{L,R}$ defined in \eqref{eq:ULR}, \eqref{eq:DLR} we 
%
find\footnote{See Appendix \ref{app:A} for the explicit formulae of $\omega^d_{ij}$ and $\rho^d_{ij}$.}
\be
({\cal D}_{L})_{ij} = \begin{cases} \omega^d_{ij} \frac{f^Q_i}{f^Q_{j}}\qquad (i<j) \\
 1\qquad \qquad(i=j)\\
 \omega^d_{ij} \frac{f^Q_j}{f^Q_{i}}\qquad (i>j)\end{cases}\,,\label{eq:DLij}
\ee
and
\be
({\cal D}_{R})_{ij} =\begin{cases} \rho^d_{ij} \frac{f^d_i}{f^d_{j}}\qquad (i<j)\\
 1\qquad\qquad(i=j)\\
\rho^d_{ij} \frac{f^d_j}{f^d_{i}}\qquad (i>j)
\end{cases}\,.
\ee
Analogous expressions hold for ${\cal U}_{L,R}$ with replacing ``$d$'' by ``$u$''.

Finally, making use of $V_\text{CKM} ={\cal U}_{L}^\dagger {\cal D}_{L}$, we obtain
\be
V_{us} = \alpha_{12}\frac{f^Q_1}{f^Q_2}\,,\qquad
V_{ub} = \alpha_{13}\frac{f^Q_1}{f^Q_3}\,,\qquad
V_{cb} = \alpha_{23}\frac{f^Q_2}{f^Q_3}\,,\label{eq:CKMFN}
\ee
with
\be
\alpha_{ij} = \sum_{k=i}^j \left(\omega^u_{ki}\right)^* \omega^d_{kj}\,.\label{eq:alphaij}
\ee

We would like to stress that the formulae given above are valid at leading order in $f^{Q,u,d}_i/f^{Q,u,d}_j$ $(i<j)$, but are exact in the entries of the 5D Yukawa couplings $\lambda^{u,d}$.

Finally, {we comment on} the complex phases in the above formulae. It can straightforwardly be seen that in general the quark masses, as given in  \eqref{eq:mbFN}--\eqref{eq:mdFN}, are complex quantities. In order to obtain positive and real values for the quark masses, the unphysical phases in  \eqref{eq:mbFN}--\eqref{eq:mdFN} have to be removed by suitable phase {redefinitions}, which will then also affect the phases of the flavour mixing matrices $\mathcal{U}_{L,R},\mathcal{D}_{L,R}$. Similarly, suitable phase redefinitions have to be performed in order to work with the standard phase convention for the CKM matrix \cite{Yao:2006px}.

\subsection{Discussion}

The formulae derived above considerably improve the widely used (see however \cite{Casagrande:2008hr}) na\"ive estimates
\be
m^{u,d}_i \sim \frac{v}{\sqrt{2}} \, \bar \lambda \,\frac{e^{kL}}{kL} f^Q_i f^{u,d}_i\,,\label{eq:naive1}
\ee
where $\bar \lambda$ is the average value of the (anarchic) 5D Yukawa couplings, and 
\be
(\mathcal{U}_L)_{ij},  (\mathcal{D}_L)_{ij} \sim \frac{f^Q_i}{f^Q_j}\,,\qquad
(\mathcal{U}_R)_{ij} \sim \frac{f^u_i}{f^u_j}\,, \qquad (\mathcal{D}_R)_{ij} \sim \frac{f^d_i}{f^d_j}\qquad (i<j)\,.\label{eq:naive2}
\ee
These estimates are obtained from assuming a completely anarchic, i.\,e. structureless, Yukawa coupling matrix. However, a random $3\times 3$ complex matrix generically does not have all entries of equal size, unless there is some symmetry enforcing such a structure. In addition the  formulae \eqref{eq:naive1} and \eqref{eq:naive2} give no hint how the complex phases present in the fundamental Yukawa couplings $\lambda^{u,d}$ are related to the effective CP-violating phases of the $\mathcal{U}_{L,R},\mathcal{D}_{L,R}$ flavour mixing matrices and to the KM phase. 

In eq.~\eqref{eq:mbFN}--\eqref{eq:CKMFN}, on the other hand, the dependence of the quark masses and mixing matrices on the elements of the fundamental Yukawa couplings $\lambda^{u,d}$ is spelled out in explicit terms, and the only approximation made is the neglect of higher order corrections in the hierarchies $f^{Q,u,d}_i/f^{Q,u,d}_j$ $(i<j)$. In general, this approximation is well justified, as the fermionic shape functions exhibit a strong hierarchy, with the weakest one in the right-handed down sector. Therefore the largest uncertainties of the above formulae are generally to be expected in $\mathcal{D}_R$. In addition, as the exact dependence on $\lambda^{u,d}$ is calculated, predictions can be made not only for the absolute size of the $\mathcal{U}_{L,R},\mathcal{D}_{L,R}$ elements, but also for their complex phases, that are relevant for CP-violation.

However  it can happen that the leading order terms given above are accidentally suppressed by the structure of the Yukawa couplings $\lambda^{u,d}$. In such a case, the na\"ively expected accuracy of the above formulae is lost, and higher order corrections have to be included in order to obtain meaningful results.

In addition, the unitarity of the flavour mixing matrices $\mathcal{U}_{L,R},\mathcal{D}_{L,R}$, and therefore also of the CKM matrix, is intrinsically violated by contributions suppressed by $f^{Q,u,d}_i/f^{Q,u,d}_j$ $(i<j)$. The exact unitarity of these matrices, however, is very important for the study of FCNC processes.

Therefore the above formulae should not be used to perform exact calculations of flavour violating observables, but are meant to give an improved estimate of the size of effects to be expected.

In our numerical analysis we will make use of the formulae \eqref{eq:mbFN}--\eqref{eq:CKMFN} in order to fit the SM quark masses and CKM mixings, which will subsequently be checked numerically. For the study of the $\Delta F=2$ observables in question we will then use the exact numerical results for the $\mathcal{D}_{L,R}$ mixing matrices.

\boldmath
\newsection{$\Delta F=2$ Transitions}\label{sec:trans}
\unboldmath
\subsection{Preliminaries}

In what follows we will use conventions and notation of~\cite{Blanke:2006sb} so that an easy comparison with the SM results and with the results obtained in the Littlest Higgs model with T-parity (LHT)  will be possible.

The SM Hamiltonians for $K^0-\bar K^0$ and $B_{s,d}^0-\bar B_{s,d}^0$ mixings, in the notation also used in the present paper, can be found in (3.1) and (3.2) of \cite{Blanke:2006sb}, respectively.
The SM contribution to the off-diagonal element $M_{12}$ in the neutral $K$and $B_{d}$ meson mass matrices {is} then given as follows
\bea\label{eq:3.4}
\left(M_{12}^K\right)_\text{SM}&=&\frac{G_F^2}{12\pi^2}F_K^2\hat
B_K m_K M_{W}^2\left[
\lambda_c^{*2}\eta_1S_c+\lambda_t^{*2}\eta_2S_t+2\lambda_c^*\lambda_t^*\eta_3S_{ct}
\right]
\,,\\
\left(M_{12}^d\right)_\text{SM}&=&{\frac{G_F^2}{12\pi^2}F_{B_d}^2\hat
B_{B_d}m_{B_d}M_{W}^2
\left[
\left(\lambda_t^{(d)*}\right)^2\eta_BS_t
\right]}\,,\label{eq:3.6}
\eea
where
 $\lambda_i=V_{is}^*V_{id}$ and  $\lambda_t^{(q)}=V_{tb}^*V_{tq}$ with $V_{ij}$ being the elements of the CKM matrix. Here, $S_c, S_t$ and $S_{ct}$ are the one-loop box functions for which explicit expressions are given e.\,g.~in \cite{Blanke:2006sb}. The factors $\eta_i$ are QCD {corrections} evaluated at the NLO level in \cite{Herrlich:1993yv,Herrlich:1995hh,Herrlich:1996vf,Buras:1990fn,Urban:1997gw}. Finally $\hat B_K$ and $\hat B_{B_d}$ are the well-known non-perturbative factors.
The amplitude $\left(M_{12}^s\right)_\text{SM}$ can be obtained
from \eqref{eq:3.6} by simply replacing $d$ by $s$.

It should be emphasised that in the SM only a single operator
\be
(\bar s d)_{V-A}(\bar s d)_{V-A} = \left[\bar s\gamma_\mu(1-\gamma_5)d\right]\otimes
\left[\bar s\gamma^\mu(1-\gamma_5)d\right]
\ee
and
\be
(\bar b q)_{V-A}(\bar b q)_{V-A}= \left[\bar b\gamma_\mu(1-\gamma_5)q\right]\otimes
\left[\bar b\gamma^\mu(1-\gamma_5)q\right]
\ee
 contributes to $M_{12}^K$ and $M_{12}^q\;(q=d,s)$, respectively. Moreover complex phases are only present in the CKM factors.

Our next goal is to generalise these formulae to include the new tree level contributions from KK gluons as shown in Fig.~\ref{fig:Figure 1}. We will see that three distinct new features will characterise these new contributions:
\begin{enumerate}
 \item The flavour structure will differ from the CKM one.
\item
FCNC transitions will appear already {at the tree level} as opposed to the one-loop SM contributions in \eqref{eq:3.4} and \eqref{eq:3.6}.
\item In addition to $(\bar s d)_{V-A}(\bar s d)_{V-A}$ and $(\bar b q)_{V-A}(\bar b q)_{V-A}$ (with $q=d,s$) new operators will be present in the effective Hamiltonians in question.
\end{enumerate}
We recall that only the first feature is present in the LHT model.
\boldmath
\subsection{Tree Level KK Gluon Contributions}
\begin{figure}
 \centering
\epsfig{file=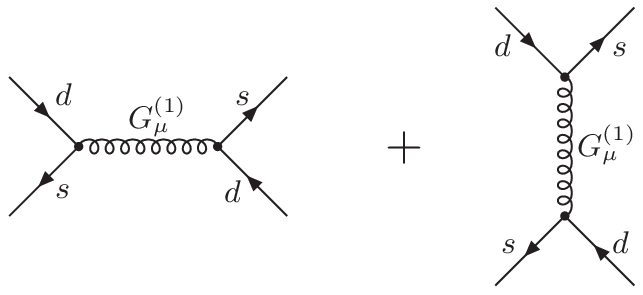,scale=1.2}
\caption{\it Tree level contribution of KK gluons to $K^0-\bar K^0$ mixing.\label{fig:Figure 1}}
\end{figure}

\unboldmath
{We begin our discussion with the tree level} exchanges of the lightest KK gluons $G_\mu^{(1)a}$ as shown in the case of $\Delta S=2$ transitions in Fig.~\ref{fig:Figure 1}.
Analogous diagrams contribute to $B_{d,s}^0-\bar B_{d,s}^0$ mixings. 
We will {analyse} tree level EW contributions {subsequently}.

The neutral current Lagrangian describing the interaction of the lightest KK gluons ${G_a^\mu}^{(1)}$ ($a=1,...,8$) with SM down quarks ($i=1,2,3$) given first in the weak eigenstate basis is flavour diagonal and given by 
\begin{equation}\label{eq:3.8}
 \mathcal{L}_\text{NC}^\text{QCD}=- p_\text{UV} \sum_i \bar\psi_i \gamma_\mu t^a \left[ \varepsilon_L(i) P_L+ \varepsilon_R(i) P_R\right] \psi_i {G_a^\mu}^{(1)}\,,
\end{equation}
where $p_\text{UV}$ parameterises the influence of brane kinetic terms on the $SU(3)_c$ coupling, as introduced in \eqref{eq:vertex}. The colour matrices $t^a$ satisfy $ \left[ t_a,t_b \right]=i f_{abc} t_c$.
We suppress the quarks' colour indices $(\alpha, \beta)$ for the moment, $P_{R,L}=(1\pm\gamma_5)/2$. The value of $p_\text{UV}$ is very important for the present analysis and is model dependent as discussed in \cite{Csaki:2008zd}. We recalled this discussion in Section \ref{sec:BKT}.

$\varepsilon_L(i)$ and $\varepsilon_R(i)$ are given by the overlap integrals of quark shape functions and the shape function of the ${G_a^\mu}^{(1)}$, with the latter strongly peaked towards the IR brane. From \eqref{eq:vertex} we have 
\be
\varepsilon_{L,R}(i) = g_s^\text{4D}\frac{1}{L}\int_0^L dy\,e^{ky} \left[f^{(0)}_{L,R}(y,c^i_\Psi)\right]^2 g(y)\,,
\label{varepsilon}
\ee
with $f^{(0)}_{L,R}(y,c^i_\Psi)$ given in \eqref{eq:fL}, \eqref{eq:fR} and $g(y)$ in \eqref{eq:gaugeprofile}.
As the shape functions of $L$ and $R$ quarks generally differ from each other, 
$\varepsilon_L(i) \neq \varepsilon_R(i)$ and parity is broken by QCD-like interactions in this model. 
Most importantly $\varepsilon_{L,R}(i)$ depend on the flavour index $i$. This dependence breaks the flavour universality of strong interactions and implies tree level FCNC transitions mediated by ${G_a^\mu}^{(1)}$ as we will see in a moment.

In what follows it will be useful to introduce two diagonal matrices 
\begin{eqnarray}\label{eq:3.10}
 \hat\varepsilon_L&=&{\rm{diag}}\left(\varepsilon_L(1), \varepsilon_L(2), \varepsilon_L(3)\right) \\
 \hat\varepsilon_R&=&{\rm{diag}}\left(\varepsilon_R(1), \varepsilon_R(2), \varepsilon_R(3)\right) 
\end{eqnarray}
and two non-diagonal matrices
\begin{eqnarray}
\hat\Delta_L&=&{\cal{D}}_L^{\dagger} \, \hat{\varepsilon}_L  \, {\cal{D}}_L\label{eq:3.11}\\
\hat\Delta_R&=&{\cal{D}}_R^{\dagger} \, \hat{\varepsilon}_R  \, {\cal{D}}_R \label{eq:3.12}
\end{eqnarray}
with ${\cal{D}}_{L,R}$ defined in \eqref{eq:DLR}. $\hat\Delta_{L,R}$
describe the FCNC couplings of down quark mass eigenstates to the lightest KK gluons. 

After {rotation to the mass eigenbasis} we find then
\be\label{eq:3.14}
 \mathcal{L}_\text{NC}^\text{QCD}\equiv-p_\text{UV} \left[ \mathcal{L}_{L}^\text{QCD}+ \mathcal{L}_{R}^\text{QCD}\right]\,,
\ee
where
\bea\label{eq:3.15}
 \mathcal{L}_{L}^\text{QCD} &=& \left[ \Delta_L^{sd}(\bar s_L \gamma_\mu t^a d_L)+\Delta_L^{bd}(\bar b_L \gamma_\mu t^a d_L) +\Delta_L^{bs}(\bar b_L \gamma_\mu t^a s_L)\right] G_a^{\mu(1)} \,,\\
\label{eq:3.16}
 \mathcal{L}_{R}^\text{QCD} &=& \left[ \Delta_R^{sd}(\bar s_R \gamma_\mu t^a d_R)+\Delta_R^{bd}(\bar b_R \gamma_\mu t^a d_R) +\Delta_R^{bs}(\bar b_R \gamma_\mu t^a s_R)\right] G_a^{\mu(1)} \,,
\eea
and $\Delta_{L,R}^{ij}$ are the elements of the matrices $\hat\Delta_{L,R}$. These elements are complex quantities and introduce new flavour and CP-violating interactions that can have a pattern very different from the CKM one.

The diagrams in Fig.~\ref{fig:Figure 1} lead to the following effective Hamiltonian for $\Delta S=2$ transitions mediated by the lightest KK gluons with mass $M_\text{KK}$:
\bea
\left[\Heff^{\Delta S=2}\right]_\text{KK}&=&\frac{{p_\text{UV}}^2}{2M_\text{KK}^2}\left[\left(\Delta_L^{sd}\right)^2\left(\bar s_L\gamma_\mu t^a d_L\right)\left(\bar s_L\gamma^\mu t^a d_L\right)\right.\nonumber\\
&& \hspace{1cm} {}
+\left(\Delta_R^{sd}\right)^2\left(\bar s_R\gamma_\mu t^a d_R\right)\left(\bar s_R\gamma^\mu t^a d_R\right)\nonumber\\
&& \hspace{1cm} {} +\left.2\Delta_L^{sd}\Delta_R^{sd}\left(\bar s_L\gamma_\mu t^a d_L\right)\left(\bar s_R\gamma^\mu t^a d_R\right)\right.\Big]\,.
\label{eq:3.17}
\eea
For the $B_{d,s}^0-\bar B_{d,s}^0$ Hamiltonians one has to replace ``$sd$'' by ``$bd$'' and ``$bs$'', respectively.

The Hamiltonian in (\ref{eq:3.17}) is valid at scales $\mathcal{O}(M_\text{KK})$ and has to be evolved to low energy scales $\mu=\mathcal{O}(2\gev)$, $\mu(m_b)$ at which the hadronic matrix elements of the operators in question can be evaluated by lattice methods. The relevant anomalous dimension matrices necessary for this renormalisation group evolution have been calculated at two-loop level in \cite{Ciuchini:1997bw,Buras:2000if} and analytic formulae for the relevant QCD factors analogous to $\eta_i$ in (\ref{eq:3.4}) and (\ref{eq:3.6}) can be found in~\cite{Buras:2001ra}. When using these formulae we neglect the unknown $\ord(\alpha_s)$ corrections to the Wilson coefficients of the relevant new operators at $\mu = M_\text{KK}$. As $\alpha_s(M_\text{KK})$ with $M_\text{KK}\simeq 3\tev$ is small, these corrections are negligible in comparison with the effects of renormalisation group running from $\mu=M_\text{KK}$ down to $\mu\sim \ord(m_b,2\gev)$.

Our next task is then to transform the operator basis in (\ref{eq:3.17}) into the basis used in~\cite{Buras:2001ra}:
\bea
\mathcal{Q}_1^{VLL}&=&\left(\bar s\gamma_\mu P_L d\right)\left(\bar s\gamma^\mu P_L d\right)\,,\nonumber\\
\mathcal{Q}_1^{VRR}&=&\left(\bar s\gamma_\mu P_R d\right)\left(\bar s\gamma^\mu P_R d\right)\,,\nonumber\\
\mathcal{Q}_1^{LR}&=&\left(\bar s\gamma_\mu P_L d\right)\left(\bar s\gamma^\mu P_R d\right)\,,\nonumber\\
\mathcal{Q}_2^{LR}&=&\left(\bar s P_L d\right)\left(\bar s P_R d\right)\,,
\label{eq:3.18}
\eea
where we suppressed colour indices as they are summed up in each factor. For instance $\bar s\gamma_\mu P_L d$ stands for $\bar s_\alpha\gamma_\mu P_L d_\alpha$ and similarly for other factors.

A straightforward calculation gives us the effective Hamiltonian\footnote{We would like to thank Michaela Albrecht for checking this result.} for $\Delta S=2$ transitions in the basis (\ref{eq:3.18}) with the Wilson coefficients corresponding to $\mu=\mathcal{O}(M_\text{KK})$ 
\bea
\left[\Heff^{\Delta S=2}\right]_\text{KK}=\frac{1}{4M_\text{KK}^2}\left[C_1^{VLL}\mathcal{Q}_1^{VLL}+C_1^{VRR}\mathcal{Q}_1^{VRR}\right.\nonumber\\
+C_1^{LR}\mathcal{Q}_1^{LR}+C_2^{LR}\mathcal{Q}_2^{LR}\big]\,,
\label{eq:3.19}
\eea
where
\bea
C_1^{VLL}(M_\text{KK})&=&\frac{2}{3}{p_\text{UV}}^2\left(\Delta_L^{sd}\right)^2\nonumber\\
C_1^{VRR}(M_\text{KK})&=&\frac{2}{3}{p_\text{UV}}^2\left(\Delta_R^{sd}\right)^2\nonumber\\
C_1^{LR}(M_\text{KK})&=&-\frac{2}{3}{p_\text{UV}}^2\Delta_L^{sd}\Delta_R^{sd}\nonumber\\
C_2^{LR}(M_\text{KK})&=&-4{p_\text{UV}}^2\Delta_L^{sd}\Delta_R^{sd}\,.
\label{eq:3.21}
\eea
Analogous expressions exist for $B_d^0-\bar B_d^0$ and $B_s^0-\bar B_s^0$ systems, with $sd$ replaced by $bd$ and $bs$, respectively. We confirm the results of \cite{Csaki:2008zd}.

\boldmath
\subsection{Tree Level Electroweak Contributions to $\Delta F=2$ Processes}
\unboldmath\label{sec:EWcont}

The KK gluon tree level contributions in Fig.~\ref{fig:Figure 1} discussed until now {are believed to} dominate the NP contributions to $\Delta F=2$ processes in the model in question. However 
{we will demonstrate now that while this  is justified 
in the case of $\eps_K$ and $\Delta M_K$, in the case of $B_{d,s}$ physics observables it is mandatory to include also tree level EW gauge boson contributions.} 
 To our knowledge the only paper studying EW contributions to $\Delta F=2$ processes is the analysis of Burdman \cite{Burdman:2002gr}. However, in that paper only $Z$ contributions have been considered, and moreover the QCD renormalisation group enhancement of {the} $\mathcal{Q}_{LR}$ operators has not been taken into account. As will be demonstrated below {in the model considered here} the dominant EW contributions do not come from $Z$ but from tree level exchanges {of the  two new heavy gauge bosons $Z_H$ and $Z'$. The contribution of the KK photon turns out to be much smaller than the latter contributions.}

Let us begin with the KK photon contribution $A^{(1)}$. The contributing diagrams are as in Fig.~\ref{fig:Figure 1} with $G^{(1)}$ replaced by $A^{(1)}$.  $\mathcal{L}_\text{NC}^\text{QED}$ is given simply by (\ref{eq:3.8}) with the colour matrices $t^a$ replaced by the identity in colour space and $\varepsilon_{L,R}(i)$ replaced by the overlap integrals similar to the ones in (\ref{varepsilon}) but with $g_s^\text{4D}$ replaced by $e^\text{4D}$, the 4D QED coupling constant. Note that the shape function of {the} KK photon  is  equal to $g(y)$. Moreover it is useful to absorb the electric charge factor in $\varepsilon_{L,R}(i)(A^{(1)})$.

The calculation is simplified relative to the previous one by the fact that in the absence of $t^a$ one immediately obtains the result in the basis (\ref{eq:3.18}).

We find the following corrections to the Wilson coefficients $C_i(M_\text{KK})$
\bea
\left[\Delta C_1^{VLL}(M_\text{KK})\right]^\text{QED}&=&2\left[\Delta_L^{sd}(A^{(1)})\right]^2\nonumber\,,\\
\left[\Delta C_1^{VRR}(M_\text{KK})\right]^\text{QED}&=&2 \left[\Delta_R^{sd}(A^{(1)})\right]^2\nonumber\,,\\
\left[\Delta C_1^{LR}(M_\text{KK})\right]^\text{QED}&=&4 \left[\Delta_L^{sd}(A^{(1)})\right]\left[\Delta_R^{sd}(A^{(1)})\right]\nonumber\,,\\
\left[\Delta C_2^{LR}(M_\text{KK})\right]^\text{QED}&=&0\,,
\label{WilsonQED}
\eea
 where $\Delta_{L,R}^{s,d}(A^{(1)})$ are analogous to $\Delta_{L,R}^{s,d}(G^{(1)})$ considered before with explicit expressions given in Appendix \ref{app:B}. The following observations should be made:

\begin{itemize}
\item $\Delta_{L,R}^{s,d}(A^{(1)})$ are suppressed by the charge factor $1/9$ and {$\alpha_\text{QED}/\alpha_s(M_\text{KK})$} relatively to $\Delta_{L,R}^{s,d}(G^{(1)})$. These suppressions are partially compensated by the absence of {the} $1/3$ colour factors in (\ref{WilsonQED}).

\item Without $\ord(\alpha_s)$ corrections to the tree level exchange of  the KK photon, the coefficient $\left[\Delta C_2^{LR}(M_\text{KK})\right]^\text{QED}$ vanishes. Strictly speaking for a NLO-QCD analysis the $\ord(\alpha_s)$ corrections to the result (\ref{WilsonQED}) should be included. But as these corrections are small we can neglect them.

\item The mixing of $\mathcal{Q}_2^{LR}$ with $\mathcal{Q}_1^{LR}$ generates through renormalisation group effects a non-vanishing {$\left[\Delta C_2^{LR}(\mu_0)\right]^\text{QED}$ that is proportional to  $\left[\Delta C_1^{LR}(M_\text{KK})\right]^\text{QED}$} and consequently is $\ord(\alpha_\text{QED})$.
\end{itemize}

We consider next the contributions of $Z$, $Z_H$ and $Z'$ gauge bosons that before EWSB correspond to the zero mode $Z^{(0)}$, its first excited KK state $Z^{(1)}$ and the heavy $Z_X^{(1)}$ gauge boson, a linear combination of $W_R^{3\mu}$ and $X_\mu$ of $U(1)_X$ \cite{Blanke:2008aa}. {Clearly while the KK gluon and photon contributions are universal to all RS models with bulk fermions, the contributions discussed in the following depend sensitively on the EW gauge group and the choice of fermion representations.}

Before EWSB the couplings of $Z^{(0)}$ to quark flavour eigenstates are universal but the ones of $Z^{(1)}$ and $Z_X^{(1)}$ are not. After EWSB the mixing between $Z^{(0)}$, $Z^{(1)}$ and $Z_X^{(1)}$ implies breakdown of universality of the couplings of the mass eigenstates $Z$, $Z_H$ and $Z'$ to quark flavour eigenstates which after the rotation to quark mass eigenstates implies tree level FCNC processes mediated by these three gauge bosons. Now the FCNC couplings of $Z_H$ and $Z'$ are $\ord(1)$, while the ones of $Z$ are $\ord(v^2/M_\text{KK}^2)$. {Consequently, its contribution to $\Delta F=2$ processes is $\ord(v^4/M_\text{KK}^4)$ and can be safely neglected already for this reason. In addition as we will demonstrate in Section \ref{sec:Z} the {flavour violating $Z$ coupling to left-handed down-type quarks} vanishes in the limit of exact $P_{LR}$ symmetry, so that these contributions are {suppressed} in the model in question also in the case of $\Delta F=1$ processes \cite{BBDGG}.}

The calculation of $\ord(v^2/M_\text{KK}^2)$ tree level contributions from $Z_H$ and $Z'$ proceeds similarly to the calculation of the $A^{(1)}$ contribution and we find: 
\bea
\left[\Delta C_1^{VLL}(M_\text{KK})\right]^\text{EW}&=&2 \left[\Big(\Delta_L^{sd}(Z^{(1)})\Big)^2+\left(\Delta_L^{sd}(Z_X^{(1)})\right)^2\right]\nonumber\,,\\
\left[\Delta C_1^{VRR}(M_\text{KK})\right]^\text{EW}&=&2 \left[\Big(\Delta_R^{sd}(Z^{(1)})\Big)^2+\left(\Delta_R^{sd}(Z_X^{(1)})\right)^2\right]\nonumber\,,\\
\left[\Delta C_1^{LR}(M_\text{KK})\right]^\text{EW}&=&4 \left[\Delta_L^{sd}(Z^{(1)})\Delta_R^{sd}(Z^{(1)})+\Delta_L^{sd}(Z_X^{(1)})\Delta_R^{sd}(Z_X^{(1)})\right]\nonumber\,,\\
\left[\Delta C_2^{LR}(M_\text{KK})\right]^\text{EW}&=&0\,,
\label{WilsonEW}
\eea
 where the overlap integrals $\Delta_{L,R}^{s,d}(Z^{(1)})$ and $\Delta_{L,R}^{s,d}(Z_X^{(1)})$ are {explicitly} given in Appendix \ref{app:B}. They include the relevant weak couplings and weak charges.

{In order to estimate the size of EW contributions when compared to the KK gluon exchanges we factor out all the couplings and charge factors from $\Delta_{L,R}^{sd}$. The remaining $\tilde{\Delta}_{L,R}^{sd}$ are then universal for all the gauge bosons considered up to the different boundary condition of $Z_X^{(1)}$ on the UV brane, whose inclusion amounts to only a percent effect on ${\Delta}_{L,R}^{sd}(Z_X^{(1)})$.

Adding {the} contributions \eqref{eq:3.21}, \eqref{WilsonQED} and \eqref{WilsonEW} and evaluating the various couplings we find
\bea
C_1^{VLL}(M_\text{KK})&=&(0.67+0.02+0.56)(\tilde\Delta_L^{sd})^2=1.25(\tilde\Delta_L^{sd})^2\nonumber\\
{C_1^{VRR}}(M_\text{KK})
&=&(0.67+0.02+0.98)(\tilde\Delta_R^{sd})^2=1.67(\tilde\Delta_R^{sd})^2\nonumber\\
C_1^{LR}(M_\text{KK})&=&(-0.67+0.04+1.13)(\tilde\Delta_L^{sd}\tilde\Delta_R^{sd})={0.50}(\tilde\Delta_L^{sd}\tilde\Delta_R^{sd})\label{WilsonEWest}
\eea 
where the three contributions correspond to KK gluon, KK photon and combined $(Z',Z_H)$ exchanges respectively.\footnote{These results are obtained neglecting the running of the EW gauge couplings between the EW scale $M_Z$ and the KK scale $M_\text{KK}$. {Taking} into account also these contributions, we would have corrections to the gauge couplings at the $5\%$ level, so that we can easily neglect them.} The Wilson coefficient $C_2^{LR}(M_\text{KK})$ receives only KK gluon contributions at $\mu=M_\text{KK}$.

We observe that the EW contributions are dominated by $Z',Z_H$ exchanges and in the case of $C_1^{VLL}$, {$C_1^{VRR}$} and $C_1^{LR}$ amount to $+87\%$, $+150\%$ and $-175\%$ corrections. In particular the sign of $C_1^{LR}(M_\text{KK})$ is reversed.

We conclude that the EW gauge boson contributions to the Wilson coefficients  $C_1^{VLL},{C_1^{VRR}}$ and $C_1^{LR}$ at $\mu=M_\text{KK}$ are of the same order as the KK gluon contributions and have to be taken into account. In the case of $\varepsilon_K$ and $\Delta M_{K}$ the strong enhancement of the coefficient $C_2^{LR}$ through QCD renormalisation group effects and the chiral enhancement of the hadronic matrix element of $\mathcal{Q}_2^{LR}$ assure that KK gluon contributions still dominate by far over EW contributions, although the reversal of the sign of $C_1^{LR}$ makes the constraints from $\varepsilon_K$ and $\Delta M_{K}$ to be slightly stronger.

However, in the case of $B_{d,s}$ physics observables the QCD renormalisation group enhancement in the LR sector is smaller than in the $K$ sector and the chiral enhancement of $\left<\mathcal{Q}_2^{LR}\right>$ and $\left<\mathcal{Q}_1^{LR}\right>$ is absent. Therefore the  $\mathcal{Q}_1^{VLL}$ operator becomes important even without the EW contributions and it is further enhanced when these contributions are taken into account.

At first sight our finding that EW contributions can compete with QCD contributions is surprising. On the other hand one should remember that KK gluon contributions similarly to EW contributions are suppressed by their large masses and the main difference between these contributions results from gauge couplings, colour factors, weak charges and renormalisation group effects. Our analysis shows that with the exception of $C_2^{LR}$ all these effects conspire to make EW heavy gauge boson contributions to be as important as the KK gluon contributions in $B_{d,s}$ physics $\Delta F=2$ observables.
}

\boldmath
\subsection{Custodial Protection of Tree Level $Z$ Contributions}
\label{sec:Z}
\unboldmath

Applying the same method already used to compute the contributions of the EW bosons $Z_H$ and $Z'$ to $\Delta F=2$ processes, we find the corrections to the Wilson coefficients due to the $Z$ exchange
\bea
\left[\Delta C_1^{VLL}(M_\text{KK})\right]^{Z}&=&2\left(\frac{M_\text{Z}}{M_\text{KK}}\right)^2 (\mathcal I_1^+)^2\left[\Delta_L^{sd}(Z^{(1)})-r\Delta_L^{sd}(Z_X^{(1)})\right]^2\,,\nonumber\\
\left[\Delta C_1^{VRR}(M_\text{KK})\right]^{Z}&=&2\left(\frac{M_\text{Z}}{M_\text{KK}}\right)^2 (\mathcal I_1^+)^2\left[\Delta_R^{sd}(Z^{(1)})-r\Delta_R^{sd}(Z_X^{(1)})\right]^2\,,\nonumber\\
\left[\Delta C_1^{LR}(M_\text{KK})\right]^{Z}&=&4\left(\frac{M_\text{Z}}{M_\text{KK}}\right)^2 (\mathcal I_1^+)^2\left[\Delta_L^{sd}(Z^{(1)})-r\Delta_L^{sd}(Z_X^{(1)})\right]\nn\\
&&\hspace{4.5cm}\cdot
\left[\Delta_R^{sd}(Z^{(1)})-r\Delta_R^{sd}(Z_X^{(1)})\right]\,,\nonumber\\
\left[\Delta C_2^{LR}(M_\text{KK})\right]^{Z}&=&0\,,
\eea 
where we have defined the quantity $r=\frac{\mathcal I_1^-}{\mathcal I_1^+}\cos\psi\cos\phi\equiv\tilde{r}\cos\psi\cos\phi${, and $\mathcal{I}_1^\pm$ are the overlaps of the gauge boson shape functions with the Higgs profile as defined in \cite{Blanke:2008aa,BBDGG}}.

We see explicitly that the $Z$ contributions to $\Delta F=2$ processes are suppressed first by $\left(M_Z/M_\text{KK}\right)^2$ relative to $Z_H$ and $Z'$ contributions. But in fact in the case of $C_1^{VLL}$ and $C_1^{LR}$ the suppression is much stronger as the custodial symmetry relevant for the  $Z b_L\bar{b}_L$ protection is also active here, and it is violated only by the boundary conditions on the UV brane. Neglecting this breakdown as in the estimates of {(\ref{WilsonEWest}) and using the couplings in Appendix \ref{app:B} \cite{Blanke:2008aa}} we find 
\bea
\Delta C_1^{VLL}(M_\text{KK})&\sim&(1-\tilde{r})^2(\tilde{\Delta}_L^{sd})^2\,,\nonumber\\
\Delta C_1^{LR}(M_\text{KK})&\sim&(1-\tilde{r})(\tilde{\Delta}_L^{sd})(\tilde{\Delta}_R^{sd})\,.
\eea
{In the limit of exact $P_{LR}$ symmetry $\tilde{r}=1$, so that these two contributions vanish. As the right-handed couplings of $Z$ are smaller anyway its contribution to $\Delta F=2$ observables is negligible.}

{We recall from the discussion in Section \ref{sec:KKfermions} that KK fermion contributions to the $Z$ couplings appear at the same order in the $v^2/M_\text{KK}^2$ expansion as the KK gauge contributions already discussed and therefore have to be considered as well. It is now important to note that provided all quark representations in the model {are} symmetric under the $P_{LR}$ exchange symmetry, as is indeed the case in the model considered,
 the custodial protection mechanism is effective also for the latter contribution. Again, small non-vanishing contributions appear due to the symmetry breaking by UV boundary conditions, but they are found numerically small as expected.
}

These findings have also implications for $\Delta F=1$ processes.
{Also
  there the left-handed couplings of $Z$ to quarks are  strongly
 suppressed by the
custodial symmetry so that new physics contributions to meson decays 
with leptons in the final state
turn out to be dominated by tree level right-handed couplings 
$Zd_R^i\bar d_R^j$  \cite{BBDGG}. This should be contrasted with the model
  considered in \cite{Casagrande:2008hr} where the protection of 
the $Zd_L^i\bar d_L^j$ couplings is absent and  tree level 
$Z$ contributions to $\Delta  F=1$ {processes}
 are expected to be significantly larger.}

\boldmath
\subsection{$M_{12}$ from KK {Gauge Bosons}}
\unboldmath
{Denoting the contributions of KK gluons to the Wilson coefficients in \eqref{eq:3.21} by $\left[C_i(M_\text{KK})\right]^G$, we finally have
\be
C_i(M_\text{KK}) = \left[C_i(M_\text{KK})\right]^G + \left[\Delta C_i(M_\text{KK})\right]^\text{QED} +\left[\Delta C_i(M_\text{KK})\right]^\text{EW}\,,
\ee
with the various contributions given in \eqref{eq:3.21}, \eqref{WilsonQED} and \eqref{WilsonEW}, respectively.}

The renormalisation group evolution from $\mu=M_\text{KK}$ to a low energy scale  $\mu_0$ can be done separately from the additive SM contribution, even if $\mathcal{Q}_1^{VLL}$ is equal up to a factor of $1/4$ to the SM operator $\left(\bar sd\right)_{V-A}\left(\bar sd\right)_{V-A}$. We recall that $\mathcal{Q}_1^{VLL}$ and $\mathcal{Q}_1^{VRR}$ renormalise without mixing with other operators and that their evolution is the same as QCD is insensitive to the sign of $\gamma_5$. But as $C_1^{VLL}(M_\text{KK})\neq C_1^{VRR}(M_\text{KK})$, their Wilson coefficients at $\mu_0$ will differ from each other. On the other hand $\mathcal{Q}_1^{LR}$ and $\mathcal{Q}_2^{LR}$ mix under renormalisation so that the RG evolution operator is a $2\times2$ matrix.

The outcome of this analysis is an effective Hamiltonian relevant at the low energy scale $\mu_0$
\bea
\left[\Heff^{\Delta S=2}\right]_\text{KK}=\frac{1}{4M_\text{KK}^2}\left[C_1^{VLL}(\mu_0)\mathcal{Q}_1^{VLL}+C_1^{VRR}(\mu_0)\mathcal{Q}_1^{VRR}\right.\nonumber\\
{} +C_1^{LR}(\mu_0)\mathcal{Q}_1^{LR}+C_2^{LR}(\mu_0)\mathcal{Q}_2^{LR}\big]\,,
\label{eq:3.22}
\eea 
with analogous expressions for {the} $\Delta B=2$ Hamiltonians.

The contribution of the KK gauge bosons $G^{(1)},A^{(1)},Z_H,Z'$ to the off-diagonal element $M_{12}^{K}$ is then obtained from
\be
2m_K\left(M_{12}^K\right)_\text{KK}^\ast=\langle\bar K^0|\left[\Heff^{\Delta S=2}\right]_\text{KK}|K^0\rangle\,.
\label{eq:3.23}
\ee
To this end one has to evaluate the hadronic matrix elements
\be
\langle\bar K^0|\mathcal{Q}_i(\mu)|K^0\rangle\equiv\langle \mathcal{Q}_i(\mu)\rangle\,.
\label{eq:3.24}
\ee 
They can be parameterised as follows
\be
\langle \mathcal{Q}_1^{VLL}(\mu)\rangle=\langle \mathcal{Q}_1^{VRR}(\mu)\rangle=\frac{2}{3}m_K^2F_K^2B_1^{VLL}(\mu)\,,
\ee
\be
\langle \mathcal{Q}_1^{LR}(\mu)\rangle=-\frac{1}{3}R(\mu)m_K^2F_K^2B_1^{LR}(\mu)\,,
\ee
\be
\langle \mathcal{Q}_2^{LR}(\mu)\rangle=\frac{1}{2}R(\mu)m_K^2F_K^2B_2^{LR}(\mu)\,,
\ee
where the $B_i$ parameters are known from lattice calculations. They are related to the parameters $B_1$, $B_5$ and $B_4$ calculated in \cite{Babich:2006bh,Becirevic:2001xt} as follows
\be
B_1^{VLL}(\mu)\equiv B_1\,,\quad B_1^{LR}(\mu)\equiv B_5\,,\quad B_2^{LR}(\mu)\equiv B_4\,,
\label{eq:3.28}
\ee
and their numerical values are given in Table \ref{tab:B_i}. It should be stressed that $B_i(\mu)$ are not renormalisation group invariant parameters in contrast to $\hat B_K$ in (\ref{eq:3.4}) but in view of the results in \cite{Buras:2001ra,Babich:2006bh,Becirevic:2001xt} it is easier to use them in this way. Finally
\be
R(\mu)=\left(\frac{m_K}{m_s(\mu)+m_d(\mu)}\right)^2\,.
\label{eq:3.29}
\ee 

Collecting all these results we find $(\mu_L=2\gev)$
\bea
\left(M_{12}^K\right)_\text{KK}=\frac{1}{12M_\text{KK}^2}m_K F_K^2\cdot\Big[\left(C_1^{VLL}(\mu_L)+C_1^{VRR}(\mu_L)\right)B_1^K\nonumber\\
-\frac{1}{2}R(\mu_L)C_1^{LR}(\mu_L)B_5^K+\frac{3}{4}R(\mu_L)C_2^{LR}(\mu_L)B_4^K\Big]^\ast\,.
\label{eq:3.30}
\eea
Analogous expressions can be derived for $\left(M_{12}^d\right)_\text{KK}$ and $\left(M_{12}^s\right)_\text{KK}$ relevant for $B_d^0-\bar B_d^0$ and $B_s^0-\bar B_s^0$ mixings, respectively. For instance $(\mu_b=4.6\gev)$
\bea
\left(M_{12}^d\right)_\text{KK}=\frac{1}{12 M_\text{KK}^2}m_{B_d}F_{B_d}^2\Big[\left(C_1^{VLL}(\mu_b)+C_1^{VRR}(\mu_b)\right)B_1^d\nonumber\\
-\frac{1}{2}R^d(\mu_b)C_1^{LR}(\mu_b)B_5^d+\frac{3}{4}R^d(\mu_b)C_2^{LR}(\mu_b)B_4^d\Big]^\ast
\label{eq:3.31}
\eea
with 
\be
R^d(\mu)=\left(\frac{m_{B_d}}{m_b(\mu)+m_d(\mu)}\right)^2\,.
\label{eq:3.32}
\ee
The values of the Wilson coefficients $C_i$ in (\ref{eq:3.31}) differ from those in (\ref{eq:3.30}) as different $\Delta^{ij}$ are involved and the scales $\mu_L$ and $\mu_b$ in (\ref{eq:3.30}) and (\ref{eq:3.31}) are different from each other. Similarly $B_i^d$ in (\ref{eq:3.31}) differ from the ones in (\ref{eq:3.30}) as now hadronic matrix elements between $B_d^0$ and $\bar B_d^0$ are evaluated.

The values for $B_i$ in the $\overline{\rm MS}$-NDR scheme that we will use in our analysis have been extracted from~\cite{Babich:2006bh} and~\cite{Becirevic:2001xt} for the $K^0-\bar K^0$ system and $B_{s,d}^0-\bar B_{s,d}^0$ system, respectively. They are collected in Table~\ref{tab:B_i}, together with the relevant values of $\mu_0$.

\begin{table}[ht]
{\renewcommand{\arraystretch}{1.3}
\begin{center}
\begin{tabular}{|c|c|c|c|c|}
\hline
&$B_1$&$B_4$&$B_5$&$\mu_0$\\
\hline
$K^0$-$\bar K^0$&0.57&0.81&0.56&2.0\gev\\
$B^0$-$\bar B^0$&0.87&1.15&1.73&4.6\gev\\
\hline
\end{tabular}
\end{center}}
\caption{\it Values of the parameters $B_i$ in the $\overline{\text{MS}}$-NDR scheme obtained in~\cite{Babich:2006bh} ($K^0$-$\bar K^0$) and~\cite{Becirevic:2001xt} ($B^0$-$\bar B^0$). The scale $\mu_0$ at which $C_i$ are evaluated is given in the last column. For $\hat B_K$ in (\ref{eq:3.4}) we use $\hat B_K=0.75\pm 0.07$~\cite{Lubicz:2008am}. 
\label{tab:B_i}}
\end{table}
\subsection{Combining SM and KK {Gauge Boson} Contributions}
The final results for $M_{12}^K$, $M_{12}^d$ and $M_{12}^s$, that govern the analysis of $\Delta F=2$ transitions in the RS model in question, are then given by
\be
M_{12}^i=\left(M_{12}^i\right)_\text{SM}+\left(M_{12}^i\right)_\text{KK}\qquad(i=K,d,s)\,,
\label{eq:3.33}
\ee
with $\left(M_{12}^i\right)_\text{SM}$ given in (\ref{eq:3.4})--(\ref{eq:3.6}) and $\left(M_{12}^i\right)_\text{KK}$ in (\ref{eq:3.30}) and (\ref{eq:3.31}).

\boldmath
\subsection{Basic Formulae for $\Delta F=2$ Observables}
\unboldmath
We collect here the formulae that we used in our numerical analysis. We would like to emphasise that, although physical observables are phase convention independent, some of the formulae collected in this section depend on the phase convention chosen for the CKM matrix  and yield correct results only if the standard phase convention {\cite{Yao:2006px}} is used consistently.

The $K_L-K_S$ mass difference is given by
\be
\Delta M_K=2\left[\RE\left(M_{12}^K\right)_\text{SM}+\RE\left(M_{12}^K\right)_\text{KK}\right]
\label{eq:3.34}
\ee
and the CP-violating parameter $\varepsilon_K$ by
\be
\varepsilon_K=\frac{\kappa_\eps e^{i\varphi_\eps}}{\sqrt{2}(\Delta M_K)_\text{exp}}\left[\IM\left(M_{12}^K\right)_\text{SM}+\IM\left(M_{12}^K\right)_\text{KK}\right]\,,
\label{eq:3.35}
\ee
where $\varphi_\eps = (43.51\pm0.05)^\circ$ and $\kappa_\eps=0.92\pm0.02$ \cite{Buras:2008nn} take into account that $\varphi_\eps\ne \pi/4$ and includes an additional effect from $\IM A_0$, the imaginary part of the 0-isospin amplitude in $K\to\pi\pi$.

For the mass differences in the $B_{d,s}^0-\bar B_{d,s}^0$ systems we have
\be
\Delta M_q=2\left|\left(M_{12}^q\right)_\text{SM}+\left(M_{12}^q\right)_\text{KK}\right|\qquad (q=d,s)\,.
\label{eq:3.36}
\ee
Let us then write \cite{Bona:2005eu}
\be
M_{12}^q=\left(M_{12}^q\right)_\text{SM}+\left(M_{12}^q\right)_\text{KK}=\left(M_{12}^q\right)_\text{SM}C_{B_q}e^{2i\varphi_{B_q}}
\label{eq:3.37}
\ee
where
\be
\left(M_{12}^d\right)_\text{SM}=\left|\left(M_{12}^d\right)_\text{SM}\right|e^{2i\beta}\,,\qquad\beta\approx 22^\circ\,,
\label{eq:3.38}
\ee
\be
\left(M_{12}^s\right)_\text{SM}=\left|\left(M_{12}^s\right)_\text{SM}\right|e^{2i\beta_s}\,,\qquad\beta_s\simeq -1^\circ\,.
\label{eq:3.39}
\ee
Here the phases $\beta$ and $\beta_s$ are defined through
\be
V_{td}=|V_{td}|e^{-i\beta}\quad\textrm{and}\quad V_{ts}=-|V_{ts}|e^{-i\beta_s}\,.
\label{eq:3.40}
\ee

We find then
\be
\Delta M_q=(\Delta M_q)_\text{SM}C_{B_q}
\label{eq:3.41}
\ee
and
\bea
S_{\psi K_S} &=& \sin(2\beta+2\varphi_{B_d})\,,
\label{eq:3.42} \\
S_{\psi\phi} &= & \sin(2|\beta_s|-2\varphi_{B_s})\,,
\label{eq:3.43}
\eea
with the latter two observables being the coefficients of $\sin(\Delta M_d t)$ and $\sin(\Delta M_s t)$ in the time dependent asymmetries in $B_d^0\to\psi K_S$ and $B_s^0\to\psi\phi$, respectively. Thus in the presence of non-vanishing $\varphi_{B_d}$ and $\varphi_{B_s}$ these two asymmetries do not measure $\beta$ and $\beta_s$ but $(\beta+\varphi_{B_d})$ and $(|\beta_s|-\varphi_{B_s})$, respectively.

At this stage a few comments on the assumptions leading to expressions (\ref{eq:3.42}) and (\ref{eq:3.43}) are in order. These simple formulae follow only if there are no weak phases in the decay amplitudes for $B_d^0\to\psi K_S$ and $B_s^0\to\psi\phi$ as is the case in the SM and also in the LHT model, where due to T-parity there are no new contributions to decay amplitudes at tree level so that these amplitudes are dominated by SM contributions. In the model discussed in the present paper new contributions to decay amplitudes with non-vanishing weak phases are present at tree level. However, these new contributions are suppressed by $M_W^2/M_\text{KK}^2$ and, as they involve charged currents, they can be safely neglected with respect to the SM tree level contributions. Basically in the present analysis we make a working assumption that tree level contributions from new physics can only be important in processes in which SM contributions are absent at tree level as is the case for $M_{12}^K$ and $M_{12}^q$ discussed above.

Now in models like the LHT model, the only operators contributing to the amplitudes $M_{12}^K$ and $M_{12}^q$ are the SM ones, that is with the $(V-A)\otimes(V-A)$ structure. Consequently the new phases $\varphi_{B_d}$ and $\varphi_{B_s}$ have purely perturbative character related to the fundamental dynamics at short distance scales.
The situation in the RS model in question is different. As now new operators contribute to the $M_{12}^q$ amplitudes the parameters $C_{B_q}$ and $\varphi_{B_q}$ in (\ref{eq:3.37}) are complicated functions of fundamental short distance parameters of the model and of the non-perturbative parameters $B_1^i$, $B_5^i$ and $B_4^i$. Thus the test of the RS model considered with the help of particle-antiparticle mixing and related CP-violation is less theoretically clean than in the case of the LHT model. On the other hand one should also emphasise that the main theoretical uncertainty in (\ref{eq:3.31}) comes from $F_{B_d}$ and not from the $B_i$ parameters.

Finally, we give the expressions for the width differences $\Delta\Gamma_q$ and the semileptonic CP-asymmetries $A_\text{SL}^q$
\be
\frac{\Delta\Gamma_q}{\Gamma_q}= -\,\left(\frac{\Delta   M_q}{\Gamma_q}\right)^\text{exp}\,\left[\text{Re}\left(\frac{\Gamma^q_{12}}{M^q_{12}}\right)^\text{SM}\frac{\cos{2\varphi_{B_q}}}{C_{B_q}}-
\text{Im}\left(\frac{\Gamma^q_{12}}{M^q_{12}}\right)^\text{SM}\frac{\sin{2\varphi_{B_q}}}{C_{B_q}}\right]\,,
\label{eq:3.44}
\ee
\be
A_\text{SL}^q=\text{Im}\left(\frac{\Gamma^q_{12}}{M^q_{12}}
\right)^\text{SM}\frac{\cos{2\varphi_{B_q}}}{C_{B_q}}-
\text{Re}\left(\frac{\Gamma^q_{12}}{M^q_{12}}\right)^\text{SM}\frac{\sin{2\varphi_{B_q}}}{C_{B_q}}\,.
\label{eq:3.45}
\ee
Theoretical predictions of both $\Delta\Gamma_q$ and $A_\text{SL}^q$ require the non-perturbative calculation of the off-diagonal matrix element $\Gamma_{12}^q$, the absorptive part of the $B_q^0-\bar B_q^0$ amplitude. We refer to Section~3.8 of~\cite{Blanke:2006sb} for further details and just quote here~\cite{Ciuchini:2003ww} 
\begin{gather}
\text{Re}\left
  (\frac{\Gamma_{12}^d}{M_{12}^d} \right)^\text{SM} = -(3.0 \pm 1.0)\cdot10^{-3}\,,\qquad
\text{Re}\left
  (\frac{\Gamma_{12}^s}{M_{12}^s} \right)^\text{SM} = -(2.6 \pm 1.0)\cdot10^{-3} \,,\label{eq:r2}\\
\text{Im}\left
  (\frac{\Gamma_{12}^d}{M_{12}^d} \right)^\text{SM} = -(6.4 \pm 1.4)\cdot
10^{-4}\,,\qquad \text{Im}\left
  (\frac{\Gamma_{12}^s}{M_{12}^s} \right)^\text{SM} = (2.6 \pm 0.5)\cdot 10^{-5}\,.\label{eq:r1}
\end{gather}

Finally, we recall the existence of a correlation between $A_\text{SL}^s$ and $S_{\psi\phi}$ that has been pointed out in~\cite{Ligeti:2006pm} and which has been investigated model-independently in \cite{Blanke:2006ig} and in the context of the LHT model in~\cite{Blanke:2006sb}. We will see below that such a correlation also exists in the model considered here.
\subsection{Summary}
In this section we have calculated the contributions of tree level KK gluon and EW gauge boson exchanges to the amplitudes $M_{12}^K$, $M_{12}^d$ and $M_{12}^s$ in  {RS models with custodial protection of the {$Z d^i_L\bar d^j_L$} coupling}. We have then given formulae for $\Delta M_K$,
$\Delta M_d$, $\Delta M_s$, $\varepsilon_K$, $S_{\psi K_S}$,
$S_{\psi\phi}$, $\Delta \Gamma_q$ and $A^q_\text{SL}$ in a form
suitable for the study of the size of the new RS contribution.
The numerical analysis of these observables will be presented in
Section~\ref{sec:num}.

While particle-antiparticle mixing in {RS models} has  already been discussed in the literature, our analysis goes beyond these papers as we performed the full renormalisation group analysis, calculated both KK gluon and EW gauge boson contributions, and also considered more observables of interest.

\newsection{Strategy for the Numerical Analysis}\label{sec:strategy}

\subsection{Flavour Parameters}

Let us begin this section by counting the flavour parameters in the quark sector, following \cite{Agashe:2004cp}, for completeness. 

First the $3\times 3$ complex 5D Yukawa coupling matrices
\be
\lambda^u\,,\qquad \lambda^d
\ee
contain each 9 real parameters and 9 complex phases. This is precisely the case of the SM.

New flavour parameters enter through the three hermitian $3\times 3$ bulk mass matrices 
\be
c_Q\,,\qquad c_u\,,\qquad c_d\,,
\ee
which bring in additional 18 real parameters and 9 complex phases.

In total  we have thus  36 real parameters and 27 complex phases at this stage. Not all of these however are physical and some of them can be eliminated by the flavour symmetry $U(3)^3$ of the 5D theory which exists in the limit of vanishing $\lambda^{u,d}$ and $c_{Q,u,d}$. {This} flavour symmetry is identical to the one present in the SM, and as in the SM 9 real parameters and 17 phases can be eliminated by making use of this symmetry. Note that one phase cannot be removed as it corresponds to the unbroken $U(1)_B$ baryon number symmetry.

We are then left with 27 real parameters and 10 complex phases to be compared to 9 real parameters and one complex phase in the SM. Evidently the new 18 real parameters and 9 phases come from the three bulk mass matrices $c_Q$, $c_u$ and $c_d$.

As already stated in Section \ref{sec:2.4}, it is convenient to work in the special basis in which the bulk mass matrices $c_{Q,u,d}$ are diagonal and thus comprise only 9 real parameters. The remaining 18 real parameters and 10 physical phases are then collected in the 5D Yukawa coupling matrices $\lambda^u$ and $\lambda^d$. For our numerical analysis it will be essential to have an efficient parameterisation of $\lambda^{u,d}$ in terms of only these parameters. Such a parameterisation  will be presented in the next section.

\subsection{A Useful Parameterisation of $\bm{\lambda^{u,d}}$ }

As every complex $3\times 3$ matrix, the 5D Yukawa matrices can always be singular value decomposed as
\be
\lambda^u = e^{i \phi_u } U_u^\dagger D_u V_u\,, \qquad 
\lambda^d = e^{i \phi_d } U_d D_d V_d\,,\label{eq:5.3}
\ee
where the $D_{u,d}$ are real and diagonal and the $U_{u,d},V_{u,d} \in  SU(3)$. The
singular value decomposed representation contains redundancies which we will
try to get rid off in the following. At this stage the {right hand sides in \eqref{eq:5.3} contain each} $(0,1)+(3,5)+(3,0)+(3,5)=(9,11)$ parameters, corresponding to 9 real parameters
and 11 phases. Two of those phases are of course spurious (see below) since
a complex $3\times3$ matrix should be described by $(9,9)$ parameters. In order to find a description in terms of physical parameters only we will use
the Euler decomposition for $SU(3)$ matrices \cite{Byrd:1997uq} 
\beq
	U(\alpha,a,\gamma,c,\beta,b,\theta,\phi) = e^{i \lambda_3 \alpha}
	e^{i \lambda_2 a}
	e^{i \lambda_3 \gamma}
	e^{i \lambda_5 c}
	e^{i \lambda_3 \beta}
	e^{i \lambda_2 b}
	e^{i \lambda_3 \theta}
	e^{i \lambda_8 \phi}
\eeq
where $a,b,c$ are mixing angles and $\alpha,\gamma,\beta,\theta,\phi$ are phases. In the basis in which $c_{Q,d,u}$ are diagonal and real we have the freedom
to make the following rephasing
\bea
 Q_L &\rightarrow& e^{i \lambda_3 \alpha_{U_d}}
	e^{-i \lambda_8 \phi_{U_u}} Q_L\\
u_R &\rightarrow& e^{-i \phi_u } 
			e^{-i \lambda_3\theta_{V_u}}
			e^{-i \lambda_8 \phi_{V_u}} u_R\\
			d_R &\rightarrow& e^{-i \phi_d }	
				e^{-i \lambda_3 \theta_{V_d}}
				e^{-i \lambda_8 \phi_{V_d}} d_R
\eea
The unitary matrices $U,V$ in a singular value decomposition are defined up to an
internal diagonal rephasing
\beq
	U D V=(U e^{i \lambda_3 A+i \lambda_8 B} ) D (e^{-i \lambda_3 A-i \lambda_8 B} V ) = U' D V',
\eeq
Using this freedom and an additional rephasing of the quark fields we find the equivalence 
\bea
\lambda^u &=&  U_u^\dagger(0,a_{U_u},\gamma_{U_u},c_{U_u},\beta_{U_u},b_{U_u},\theta_{U_u},0)
\,D_u\, 
V_u(\alpha_{V_u},a_{V_u},\gamma_{V_u},c_{V_u},\beta_{V_u},b_{V_u},0,0)
\nn\\
&=& U_u^\dagger(0,a_{U_u},\gamma_{U_u}+r,c_{U_u},\beta_{U_u}-r,b_{U_u},\theta_{U_u},r/\sqrt{3})  \,
D_u \nn\\
&&\hspace{4cm}
V_u(\alpha_{V_u},a_{V_u},\gamma_{V_u}+r,c_{V_u},\beta_{V_u}- r,b_{V_u},0,r/\sqrt{3}) \,.
\eea
The entries $r/\sqrt{3}$ can be again rotated to zero due to the freedom to rephase
the quark zero modes. Using this invariance parameterised by $r$ allows us to choose $\gamma_{U_u} =0 $. We can now define $\lambda^u$ and $\lambda^d$ in terms of physical parameters only
\bea
\lambda^u &=&  U_u^\dagger(0,a_{U_u},0,c_{U_u},\beta_{U_u},b_{U_u},\theta_{U_u},0)  
\,D_u\, 
V_u(\alpha_{V_u},a_{V_u},\gamma_{V_u},c_{V_u},\beta_{V_u},b_{V_u},0,0)\,,\\
\lambda^d &=&  U_d(0,a_{U_d},\gamma_{U_d},c_{U_d},\beta_{U_d},b_{U_d},0,0) \, D_d\, V_d(\alpha_{V_d},a_{V_d},\gamma_{V_d},c_{V_d},\beta_{V_d},b_{V_d},0,0)\,,\qquad
\eea
with $D_u=\diag (y_u^1,y_u^2,y_u^3)$ and $D_d=\diag (y_d^1,y_d^2,y_d^3)$.
Altogether we find $18$ real parameters and $10$ physical phases contained in the 5D Yukawas, as expected. 


{

\boldmath
\subsection{Guideline for the Parameter Scan}
\unboldmath

The starting point of our numerical analysis is the generation of random 5D Yukawa coupling matrices $\lambda^{u,d}$. This can efficiently be done by means of the parameterisation presented in the previous section. In our scan we take
\be
0\le y^i_{u,d} \le 3\qquad (i=1,2,3)\,,
\ee
where the upper bound stems from the perturbativity constraint on $\lambda^{u,d}$. The mixing angles
\be
a_{U_u},c_{U_u},b_{U_u},a_{V_u},c_{V_u},b_{V_u},
a_{U_d},c_{U_d},b_{U_d},a_{V_d},c_{V_d},b_{V_d}
\ee
and CP-violating phases
\be
\beta_{U_u},\theta_{U_u},\alpha_{V_u},\gamma_{V_u},\beta_{V_u},\gamma_{U_d},\beta_{U_d},\alpha_{V_d},\gamma_{V_d},\beta_{V_d},
\ee
will be varied in their physical ranges $[0,\pi/2]$ and $[0,2\pi]$, respectively. 

The overall scale for the bulk mass parameters $c_{Q,u,d}^i$ will be fixed by the requirement that $c_Q^3$ lies in the range
\be
0.1\le c_Q^3 \le 0.5\,,
\ee
allowing for consistency with EW precision data thanks to the protection of the $Zb_L\bar b_L$ coupling {\cite{Agashe:2006at,Carena:2007ua}}. The remaining bulk mass parameters will then be fitted making use of the analytic formulae of Section~\ref{sec:FN}.

Having generated a 5D parameter point, we check its consistency with the measured quark masses and CKM parameters by diagonalising numerically the obtained effective 4D Yukawa coupling matrices $Y_{u,d}$. Note that in order not to depend on unphysical phases at this stage, we choose to fit the Jarlskog determinant \cite{Jarlskog:1988ii}
\be
J_\text{CP}=\IM(V_{ud}^{}V_{cs}^{}V_{us}^*V_{cd}^*)
\ee
rather than the CKM angle $\gamma=\text{arg}(V_{ub})$. 

Finally, in order to be able to work with the well-known formulae summarised in Section \ref{sec:trans}, we remove unphysical phases by proper phase redefinitions of the quark fields, requiring real and positive masses and standard CKM phase conventions.

Throughout the major part of our analysis, we will keep fixed 
\be
f=1\tev\,,
\ee
corresponding to
\be
M_\text{KK}\simeq2.45\tev\,.
\ee
Only in the last part of our analysis, where we wish to determine a generic bound on the KK scale by the requirement of naturalness, we will also vary  $f$.

}

\newsection{Numerical Analysis}
\label{sec:num}

\subsection{Introduction}

Having at hand all the relevant formulae for $\Delta F=2$ processes in the RS model 
in question, we will investigate how much fine-tuning of parameters is necessary in order
to obtain a satisfactory description of the existing data and whether some characteristic 
patterns of deviations from MFV can be attributed to this model.

To this end we will use the  measure of fine-tuning introduced by Barbieri and Giudice \cite{Barbieri:1987fn} and most commonly used in the literature.
In that paper the amount of tuning $\Delta_\text{BG}(O_i,p_j)$ in an observable $O_i$ with respect to a parameter $p_j$ is defined as the sensitivity of $O_i$ to infinitesimal variations of $p_j$. Explicitly,
\be
\Delta_\text{BG} (O_i,p_j) = \left| \frac{p_j}{O_i}\frac{\partial O_i}{\partial p_j}\right| \,,
\ee
where the normalisation factor $p_j/O_i$ appears in order not to be sensitive to the absolute size of $p_j$ and $O_i$. The overall fine-tuning in the observable $O_i$ is then given by
\be
\Delta_\text{BG}(O_i)= \text{max}_{j=1,\dots,m} \{  \Delta_\text{BG}(O_i,p_j) \}\,,
\ee
where the index $j$ runs over all $m$ dimensions of parameter space. Obviously, the larger $\Delta_\text{BG}(O_i)$, the more sensitive is the value of $O_i$ to small variations in the parameters $p_j$, i.\,e. the more fine-tuning is required to keep $O_i$ stable.

Now, the {RS model} in question has many parameters. Moreover $\Delta M_d$ and  $\Delta M_s$ suffer
from sizable uncertainties originating dominantly in $F_{B_s}$ and $F_{B_d}$ that enter squared
in $\Delta M_{d,s}$ and are known only within 10\% accuracy. This should be contrasted with 
the CP-violating parameter $\eps_K$, where the decay constant $F_K$ is known with 1\% 
accuracy and the parameter $\hat{B}_K$, that enters $\eps_K$ linearly, should be known within 
3\% accuracy already at the end of 2008 from lattice calculations with dynamical fermions {\cite{Antonio:2007pb}}. 
Finally let us recall that the CP-asymmetries $S_{\psi\phi}$ and $S_{\psi K_S}$ are basically 
free from hadronic uncertainties and the hadronic uncertainties in the {ratio}
$\Delta M_d/\Delta M_s$ amount to roughly 4\%.

On the experimental side the data on $\Delta M_d$,  $\Delta M_s$ and $\eps_K$ are very 
precise, so that their experimental errors can be neglected for all practical purposes, while
$S_{\psi K_S}$ is known with an uncertainty of $\pm 4\%$. $\Delta M_K$, while very accurately 
measured, is subject to poorly known long distance contributions
 and we will only require that  
$(\Delta M_K)_{\text{exp}}$ is reproduced within $\pm 50\%$.

\begin{table}[ht]
\renewcommand{\arraystretch}{1}\setlength{\arraycolsep}{1pt}
\center{\begin{tabular}{|l|l|}
\hline
$\lambda=|V_{us}|= 0.226(2)$ & $G_F= 1.16637\cdot 10^{-5}\gev^{-2}$ \qquad {} \\
$|V_{ub}| = 3.8(4)\cdot 10^{-3}$ &  $M_W = 80.403(29) \gev$ \\
$|V_{cb}|= 4.1(1)\cdot 10^{-2}$ \hfill\cite{Bona:2006sa}& $\alpha(M_Z) = 1/127.9$ \\\cline{1-1}
$\gamma = 80(20)^\circ $ & $\sin^2\theta_W = 0.23122$\\\cline{1-1}
$\Delta M_K= 0.5292(9)\cdot 10^{-2} \,\text{ps}^{-1}$ \qquad {} & $m_K^0= 497.648\mev$ \\
$|\eps_K|= 2.232(7)\cdot 10^{-3}$ \hfill\cite{Yao:2006px}& $m_{B_d}= 5279.5\mev$ \\\cline{1-1}
$\Delta M_d = 0.507(5) \,\text{ps}^{-1}$ & $m_{B_s} = 5366.4\mev$ \hfill\cite{Yao:2006px} \\\cline{2-2}
$\Delta M_s = 17.77(12) \,\text{ps}^{-1}$  & $\eta_1= 1.32(32)$ \hfill\cite{Herrlich:1993yv}\\\cline{2-2}
$S_{\psi K_S}= 0.681(25)$ \hfill\cite{Barberio:2007cr}&  $\eta_3=0.47(5)$ \hfill\cite{Herrlich:1995hh,Herrlich:1996vf} \\\hline
$\bar m_c = 1.30(5)\gev$ & $\eta_2=0.57(1)$ \\
$\bar m_t = 162.7(13)\gev$ & $\eta_B=0.55(1)$ \hfill \cite{Buras:1990fn,Urban:1997gw} \\\hline
$F_K = 156(1)\mev$ \hfill \cite{Flavianet}& $F_{B_s} = 245(25)\mev$ \\\cline{1-1}
$\hat B_K= 0.75(7)$ & $F_{B_d} = 200(20)\mev$ \\
$\hat B_{B_s} = 1.22(12)$ & $F_{B_s} \sqrt{\hat B_{B_s}} = 270(30)\mev$ \\
$\hat B_{B_d} = 1.22(12)$ & $F_{B_d} \sqrt{\hat B_{B_d}} = 225(25)\mev$ \\
$\hat B_{B_s}/\hat B_{B_d} = 1.00(3)$ \hfill \cite{Lubicz:2008am}& $\xi = 1.21(4)$ \hfill \cite{Lubicz:2008am}
 \\\hline
{$\alpha_s(M_Z)=0.118(2)$} & \\\hline
\end{tabular}  }
\caption {\textit{Values of the experimental and theoretical
    quantities used as input parameters.} 
\label{tab:input}}
\renewcommand{\arraystretch}{1.0}
\end{table}

\begin{table}[ht]
\renewcommand{\arraystretch}{1}\setlength{\arraycolsep}{1pt}
\center{\begin{tabular}{|c|cccc|}
\hline
 & $\mu=2\gev$ & $\mu = 4.6\gev$ & $\mu=172\gev$ & $\mu=3\tev$ \\\hline
$m_u(\mu)$ & $3.0(10)\mev$ & $2.5(8)\mev$ & $1.6(5)\mev$ & $1.4(5)\mev$ \\ 
$m_d(\mu)$ & $6.0(15)\mev$ & $4.9(12)\mev$ & $3.2(8)\mev$ & $2.7(7)\mev$ \\
$m_s(\mu)$ & $110(15)\mev$ & $90(12)\mev$ & $60(8)\mev$ & $50(7)\mev$ \\
$m_c(\mu)$ & $1.04(8)\gev$ & $0.85(7)\gev$ & $0.55(4)\gev$ & $0.45(4)\gev$ \\
$m_b(\mu)$ & --- & $4.2(1)\gev$ & $2.7(1)\gev$ & $2.2(1)\gev$ \\
$m_t(\mu)$ & --- & ---& $162(2)\gev$ & $135(2)\gev$
 \\\hline
\end{tabular}  }
\caption{\label{tab:masses}\it 
Renormalised quark masses at various scales, evaluated using NLO running. The $1\sigma$ uncertainties are given in brackets.}
\renewcommand{\arraystretch}{1.0}
\end{table}

{With this pattern} of uncertainties in mind, we will perform our numerical analysis in several steps
as follows:

\subsubsection*{Step 1}

We will require that the masses of the SM quarks are reproduced within 
$2\sigma$.
For the three mixing angles of the CKM matrix represented in our analysis by $|V_{us}|$,  
$|V_{ub}|$ and $|V_{cb}|$, we will require agreement within $2\sigma$. As the value of the phase
$\gamma=\delta_\text{CKM}$ from tree level decays still suffers from large uncertainties, we will 
just require that it lies in the range $60^{\circ} \leq \gamma \leq 100^{\circ}$.
The strategy for performing efficiently step 1 has been outlined in the previous section.  

\subsubsection*{Step 2}

Having {constrained} moderately the space of parameters in the first step we will investigate how much fine-tuning is necessary in order to reproduce the experimental value of $\varepsilon_K$. Similarly, we will consider the cases of $\Delta M_K$, being sensitive to $\RE(M^K_{12})$, and $S_{\psi K_S}$, being the most accurately known $\Delta F=2$ observable in the $B$ systems. 

\subsubsection*{Step 3}

At this stage we will impose the experimental constraints from $\Delta F=2$ observables. In order not to complicate our analysis, we will set all input parameters collected in Table \ref{tab:input} to their central values and instead allow the resulting observables $\Delta M_K$, $\Delta M_d$, $\Delta M_s$, $\Delta M_d/\Delta M_s$, $\eps_K$ and $S_{\psi K_S}$ to {deviate by} $\pm 50\%$, $\pm 30\%$, $\pm 30\%$, $\pm 20\%$,  $\pm 30\%$ and $\pm 20\%$, respectively. These uncertainties may appear rather conservative, but we do not want to miss any interesting effect by imposing too optimistic constraints. A similar strategy for the error analysis has been followed in \cite{Blanke:2006sb,Blanke:2006eb} in the context of the LHT model. Recently that analysis has been updated and extended by a more careful error analysis \cite{Blanke:2008ac}, revealing that the simplified error analysis of \cite{Blanke:2006sb,Blanke:2006eb} did not have a qualitative impact on the results obtained.

Furthermore additional theoretical uncertainties enter our analysis due to the several approximations made. First of all we are taking into account only the first {gauge KK modes} and not the full {KK towers}, whose neglect amounts to an error $\lsim 10\%$, which we have checked numerically. However one should keep in mind that the model becomes non-perturbative already after the first few KK modes, so that we think the result obtained from the sum over the whole KK tower cannot be fully trusted. In addition we do not numerically include the effect of mixing of the SM quarks with their heavy KK partners, which turns out to be subleading and at the level of 10\% (see Section \ref{sec:KKfermions} for details).

\subsubsection*{Step 4}

Having at hand those regions of the parameter space that are consistent with all available constraints, we study the {KK gauge boson effects} on those $\Delta F=2$ observables that are not yet known with good accuracy. These are mostly the CP-asymmetries $S_{\psi\phi}$ and $A^s_\text{SL}$, but also the width difference $\Delta\Gamma_s/\Gamma_s$, in the $B_s$ system.

\subsubsection*{Step 5}

Finally we will investigate whether the results obtained in Step 4 depend significantly on the fine-tuning $\Delta_\text{BG}(\eps_K)$. Therefore we will impose the additional constraint  $\Delta_\text{BG}(\eps_K)<20$ and redo the phenomenological analysis performed in Step 4.

{
\subsubsection*{Step 6}

Last but not least, motivated by the analysis in \cite{Csaki:2008zd}, we will derive a generic lower bound on $M_\text{KK}$ from $\eps_K$, demanding that the average required fine-tuning to get an acceptable $\eps_K$ value does not exceed a certain naturalness limit. In this step we will therefore also vary the scale $f$.

}

\vspace{4mm}

\noindent Throughout our analysis we will consider density plots rather than scatter plots, as these offer the additional information which effects are the most likely ones. We also show the colour bar for each of the plots, although the absolute number of points in each counting bin depends of course on the number of points considered and on the bin size chosen.

\subsection{Results}

\boldmath
\subsubsection{{RS} Contribution to $M^{K,d,s}_{12}$}
\unboldmath

\begin{figure}
\begin{center}
\begin{minipage}{7cm}
\epsfig{file=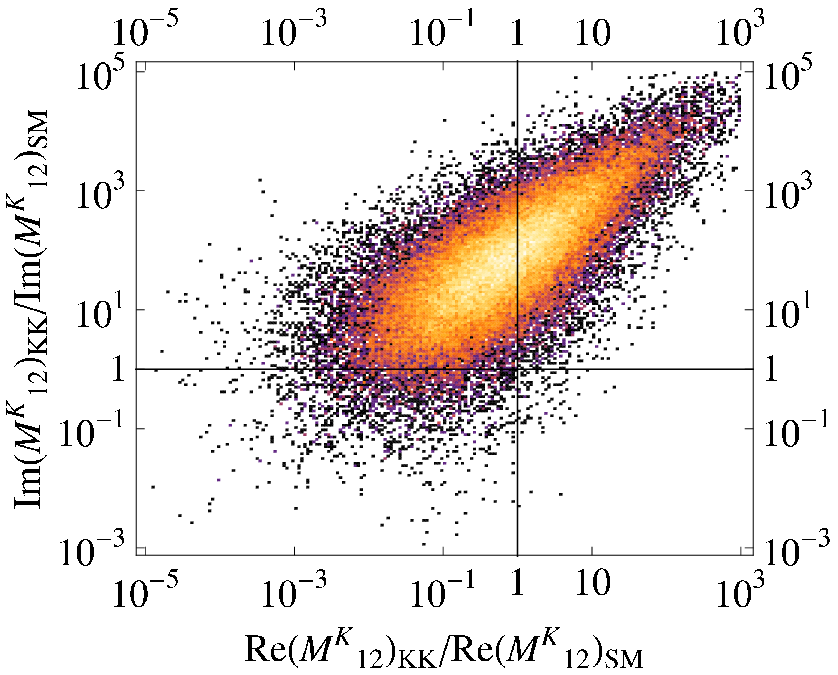,scale=.75}
\end{minipage}
\begin{minipage}{7cm}
\epsfig{file=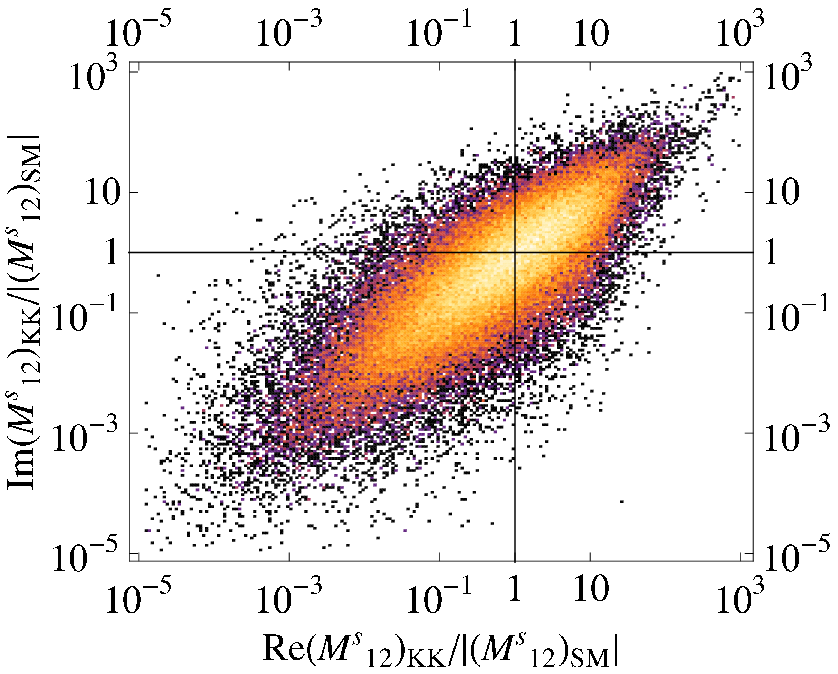,scale=.75}
\end{minipage}
\begin{minipage}{1cm}
\epsfig{file=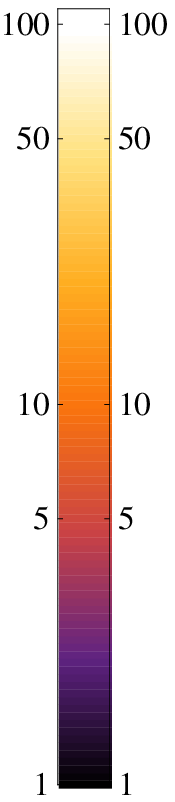,scale=.6}
\end{minipage}
\end{center}
\caption{\label{fig:MKs12}  \it left:  $\RE(M^K_{12})_\text{KK}/\RE(M^K_{12})_\text{SM}$ {and}
$\IM(M^K_{12})_\text{KK}/\IM(M^K_{12})_\text{SM}$, plotted on logarithmic axes.
right: $\RE(M^s_{12})_\text{KK}$ and $\IM(M^s_{12})_\text{KK}$, normalised to $|(M^s_{12})_\text{SM}|$ and plotted on logarithmic axes.
}
\end{figure}

In order to get a feeling for the size of the {RS} contribution to $\Delta F=2$ observables, we show in Fig.~\ref{fig:MKs12} the complex $(M^{K}_{12})_\text{KK}$ and $(M^{s}_{12})_\text{KK}$ planes.

In the left panel of Fig.~\ref{fig:MKs12}, we show 
$\IM(M^K_{12})_\text{KK}/\IM(M^K_{12})_\text{SM}$ plotted as a function of
 $\RE(M^K_{12})_\text{KK}/\RE(M^K_{12})_\text{SM}$. We observe that while $\RE(M^K_{12})_\text{KK}$ has the tendency to be generically somewhat smaller, albeit still competitive, with the SM contribution, the KK contribution to  $\IM(M^K_{12})$ typically exceeds the SM by {two orders} of magnitude. This is due to the suppression of $\IM(M^K_{12})_\text{SM}$ with respect to $\RE(M^K_{12})_\text{SM}$ by roughly a factor 100, and leads to the generic strong constraint from $\eps_K$ on the KK scale identified in \cite{Csaki:2008zd}. Still, already from this figure we can deduce that there exist regions of the parameter space for which $\IM(M^K_{12})_\text{KK}\lsim \IM(M^K_{12})_\text{SM}$, so that agreement with the data on $\eps_K$ can be obtained even for a scale as low as $M_\text{KK}= 2.5\tev$.

In the right panel of Fig.~\ref{fig:MKs12}, we show $\RE(M^s_{12})_\text{KK}$ and $\IM(M^s_{12})_\text{KK}$, normalised to $|(M^s_{12})_\text{SM}|$. {We observe that the KK gauge boson contribution tends to be of roughly the same size as the SM contribution,} and that contrary to the SM $\RE(M^s_{12})_\text{KK}$ and $\IM(M^s_{12})_\text{KK}$ are generically of the same size, so that an $\ord(1)$ new physics phase can be expected.
For completeness we mention that the case of  $M^d_{12}$ is very similar to $M^s_{12}$, and we do not show it here.

\begin{figure}
\begin{center}
\begin{minipage}{7cm}
\epsfig{file=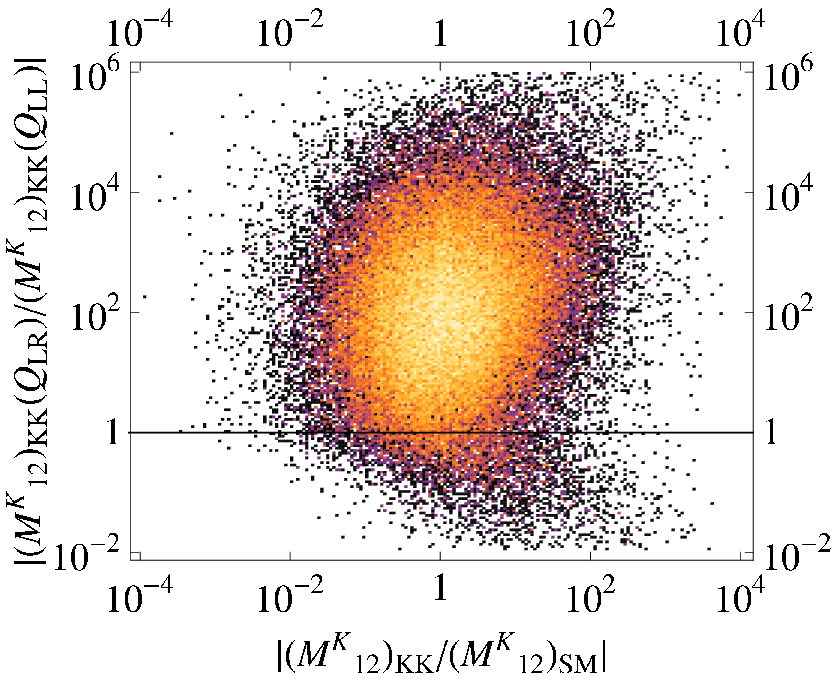,scale=.75}
\end{minipage}
\begin{minipage}{7cm}
\epsfig{file=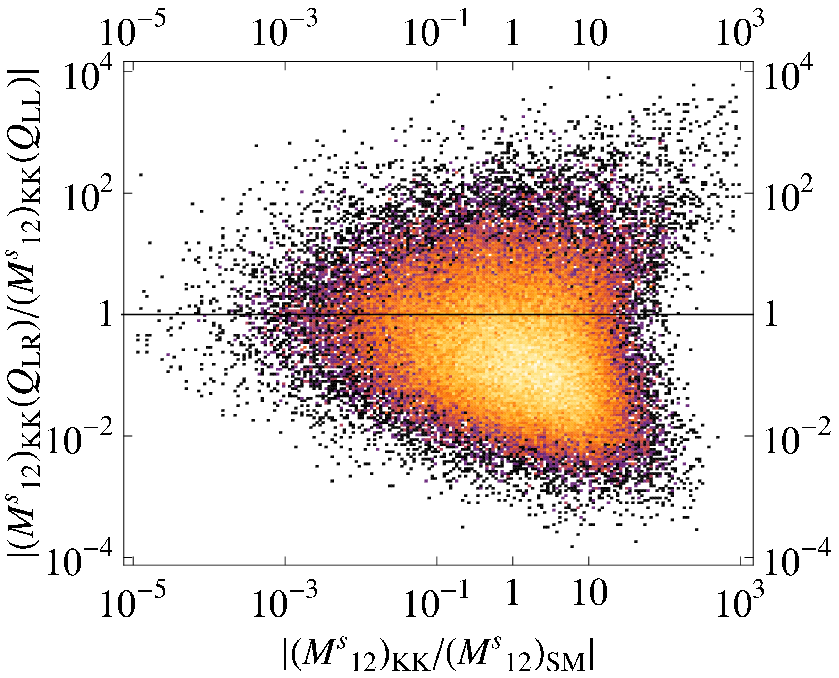,scale=.75}
\end{minipage}
\begin{minipage}{1cm}
\epsfig{file=bar100.eps,scale=.6}
\end{minipage}
\end{center}
\caption{\label{fig:LL-LR} \it The ratio of the contribution of only {$\mathcal{Q}_{LR}$ and only $\mathcal{Q}_{LL}$} to $(M^K_{12})_\text{KK}$ (left) and $(M^s_{12})_\text{KK}$ (right), as a function of $(M^i_{12})_\text{KK}/(M^i_{12})_\text{SM}$ ($i=K,s$).}
\end{figure}

Next, we aim to analyse the importance of the various operators induced by the KK {gauge boson} exchange. Therefore in the left panel of Fig.~\ref{fig:LL-LR} we show the ratio of the {$\mathcal{Q}_{LR}$ and $\mathcal{Q}_{LL}$} operator contributions to $(M^K_{12})_\text{KK}$. In accordance with the analysis in \cite{Csaki:2008zd} we observe that the LR contribution is by far the dominant one, while the LL  contribution is typically below 10\%. The reasons for this dominance is the chiral and QCD enhancement of the LR operator. The contribution of $\mathcal{Q}_{RR}$ turns out to be negligibly small, which is due to the fact that {the right-handed bulk mass parameters $c_d^i$
violate the flavour symmetry much less strongly than the left-handed $c_Q^i$ ones. In addition the $b_R$ quark 
 lives closer to the UV brane than the $b_L$ one and is therefore much less sensitive to the flavour violation induced by KK modes close to the IR brane.}

In the right panel of Fig.~\ref{fig:LL-LR} then the ratio of the {$\mathcal{Q}_{LR}$ and $\mathcal{Q}_{LL}$} operator contributions to $(M^s_{12})_\text{KK}$ is shown.
In that case the situation differs from the $K$ system, due to absence of the chiral enhancement and the weaker renormalisation group QCD enhancement. Indeed we find that the $\mathcal{Q}_{LL}$ and $\mathcal{Q}_{LR}$ turn out to be competitive in size, and in most cases $\mathcal{Q}_{LL}$ even yields the dominant contribution. {We note that while the $\mathcal{Q}_{LR}$ contribution is essentially unaffected by the EW contributions, they enhance $\mathcal{Q}_{LL}$ by roughly a factor 2, so that the importance of LL contributions in $B_{d,s}$ physics is increased by these contributions.} Again the contribution from $\mathcal{Q}_{RR}$ to $M^s_{12}$ is very small. The situation in the $B_d$ system is very similar and we do not show it explicitly.

\boldmath
\subsubsection{Fine-Tuning in $\Delta F=2$ Observables}
\unboldmath

\begin{figure}
\begin{center}
\begin{minipage}{7cm}
\epsfig{file=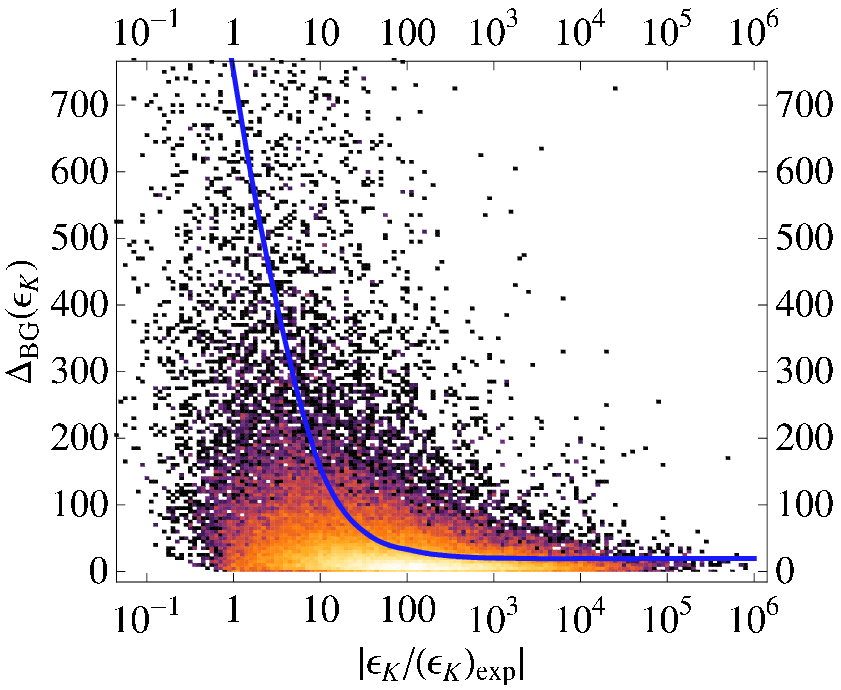,scale=.75}
\end{minipage}
\begin{minipage}{7cm}
\epsfig{file=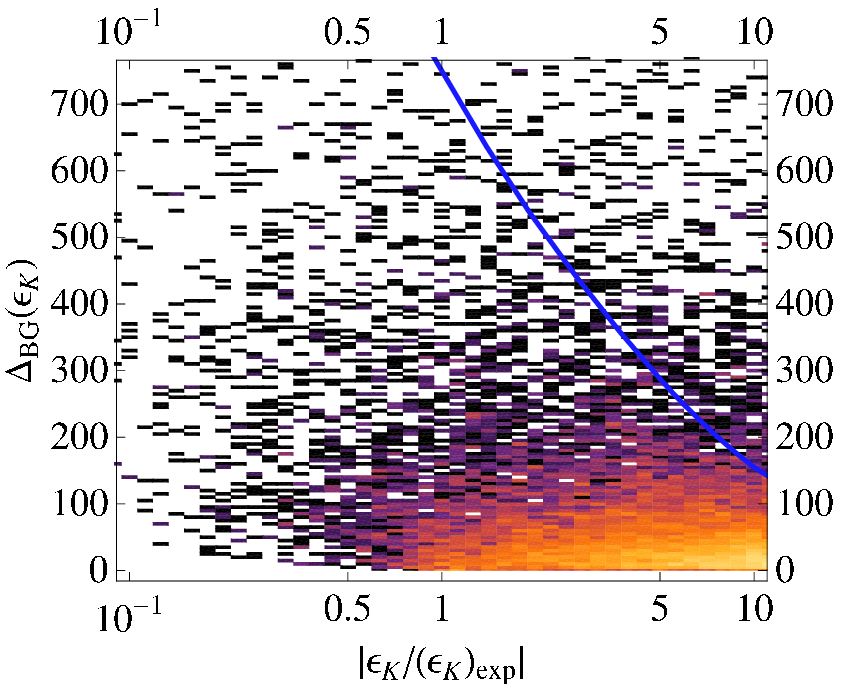,scale=.75}
\end{minipage}
\begin{minipage}{1cm}
\epsfig{file=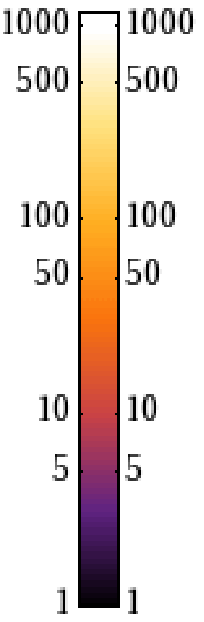,scale=.6}
\end{minipage}
\end{center}
\caption{\label{fig:epsK-tuning}\it left: 
{The fine-tuning $\Delta_\text{BG}(\eps_K)$ plotted against
$\eps_K$, normalised to its experimental value}. The blue line displays the average fine-tuning as a function of $\eps_K$. right: The same, but displaying only the phenomenologically interesting region $0.1 < |\eps_K/(\eps_K)_\text{exp}|<10$.
}
\end{figure}

While deducing already from Fig.~\ref{fig:MKs12} the possibility to obtain $\eps_K$ in accordance with the data, we are now interested in how natural such values are. Therefore we show in Fig.~\ref{fig:epsK-tuning} the fine-tuning in $\eps_K$, $\Delta_\text{BG}(\eps_K)$, as a function of $\eps_K$. {We observe that while for generic values $\eps_K/(\eps_K)_\text{exp}\sim\ord(100)$, the fine-tuning is typically relatively small, $\Delta_\text{BG}(\eps_K)\sim 20$, the average required tuning strongly increases with decreasing $\eps_K$, so that generically for $\eps_K\sim(\eps_K)_\text{exp}$ a fine-tuning of the order $\Delta_\text{BG}(\eps_K)\sim {700}$ is required, i.\,e. the amount of fine-tuning increases by roughly a factor $30-40$ when going from the generic prediction for $\eps_K$ down to values in accordance with experiment. In other words, a relative fine-tuning at the few percent level {is} on average required in order to obtain $\eps_K\sim(\eps_K)_\text{exp}$.} However, it can also be observed that although for smaller values of $\eps_K$ large fine-tunings become more likely, even for SM-like $\eps_K$ roughly 30\% of the points lie still in the range with small tuning, $\Delta_\text{BG}(\eps_K) < 20$, so that fully natural solutions to the ``$\eps_K$ problem'' can be identified.

Let us next consider the necessary amount of fine-tuning in other $\Delta F=2$ observables. As examples we show here $\Delta M_K$, being sensitive to $\RE(M^K_{12})$, and $S_{\psi K_S}$, being the most accurately known $\Delta F=2$ observable in the $B$ systems.

\begin{figure}
\begin{center}
\begin{minipage}{9cm}
\epsfig{file=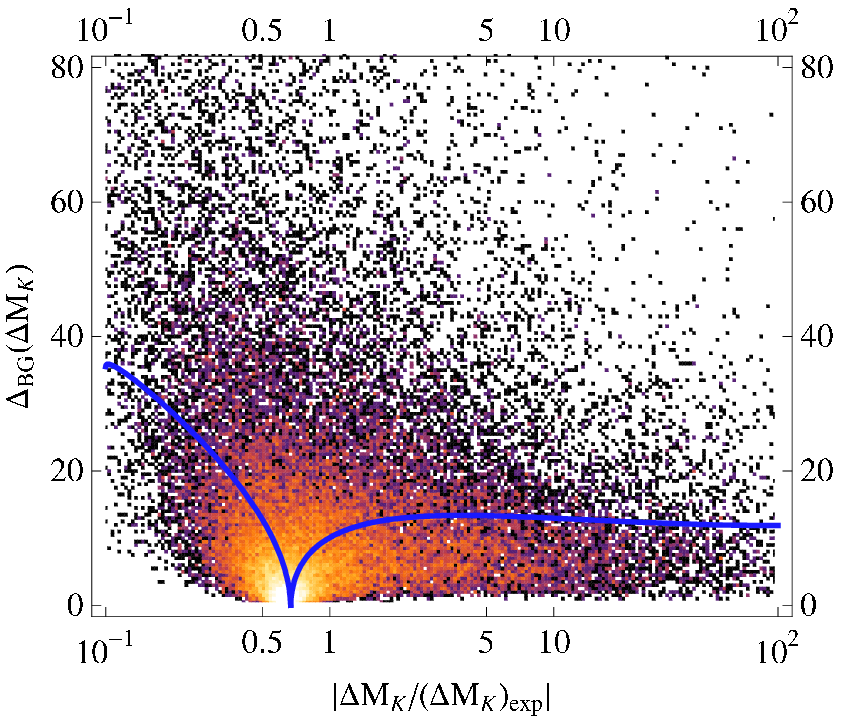,scale=1}
\end{minipage}
\begin{minipage}{1.5cm}
\epsfig{file=bar100.eps,scale=.7}
\end{minipage}
\end{center}
\caption{\label{fig:DMK-tuning}\it {The fine-tuning $\Delta_\text{BG}(\Delta M_K)$ plotted against $\Delta M_K$, normalised to its experimental value.}
The blue line displays the average fine-tuning as a function of $\Delta M_K$.}
\end{figure}

In Fig.~\ref{fig:DMK-tuning} we show the fine-tuning $\Delta_\text{BG}(\Delta M_K)$ as a function of $\Delta M_K$. We observe that the tuning is generally smaller ($\Delta_\text{BG}(\Delta M_K)\lsim 20$) than in the case of $\eps_K$, and that the smallest average values are obtained for $\Delta M_K$ in accordance with experiment. This could already be expected from Fig.~\ref{fig:MKs12}, where we found the KK contribution to $\RE(M^K_{12})$ to be of the same order of magnitude as the SM contribution, so that generically $\Delta M_K\sim (\Delta M_K)_\text{exp}$.

\begin{figure}
\begin{center}
\begin{minipage}{9cm}
\epsfig{file=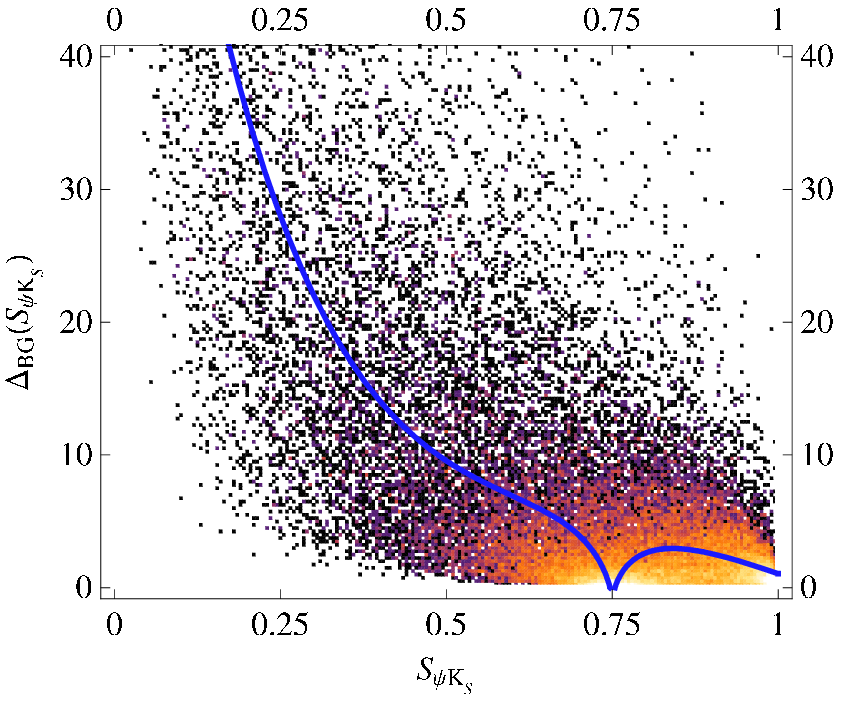,scale=1}
\end{minipage}
\begin{minipage}{1.5cm}
\epsfig{file=bar100.eps,scale=.7}
\end{minipage}
\end{center}
\caption{\label{fig:SpsiKS-tuning}\it {The fine-tuning $\Delta_\text{BG}(S_{\psi K_S})$ plotted against
$S_{\psi K_S}$.} The blue line displays the average fine-tuning as a function of $S_{\psi K_S}$.}
\end{figure}

In Fig.~\ref{fig:SpsiKS-tuning} we show $\Delta_\text{BG}(S_{\psi K_S})$ as a function of $S_{\psi K_S}$. Also in that case the average fine-tuning is smallest for $S_{\psi K_S}$ in accordance with the data. In addition the overall scale of $\Delta_\text{BG}(S_{\psi K_S})$ turns out to be different, so that typically $\Delta_\text{BG}(S_{\psi K_S})\lsim 5$. 

\boldmath
\subsubsection{Full $\Delta F=2$ Analysis  and CP-Violation in $B_s-\bar B_s$ Mixing}
\unboldmath

\begin{figure}
\begin{center}
\begin{minipage}{7cm}
\epsfig{file=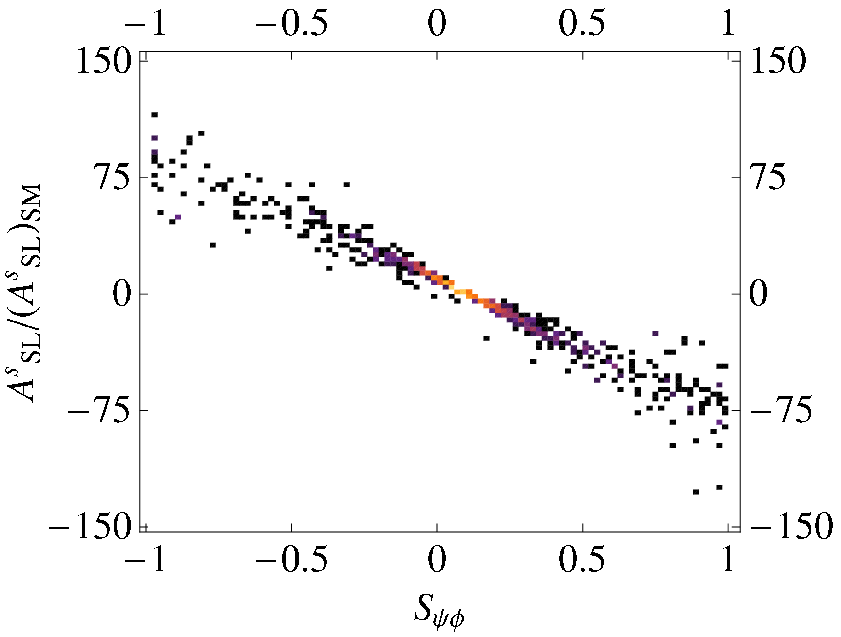,scale=.75}
\end{minipage}
\begin{minipage}{7cm}
\epsfig{file=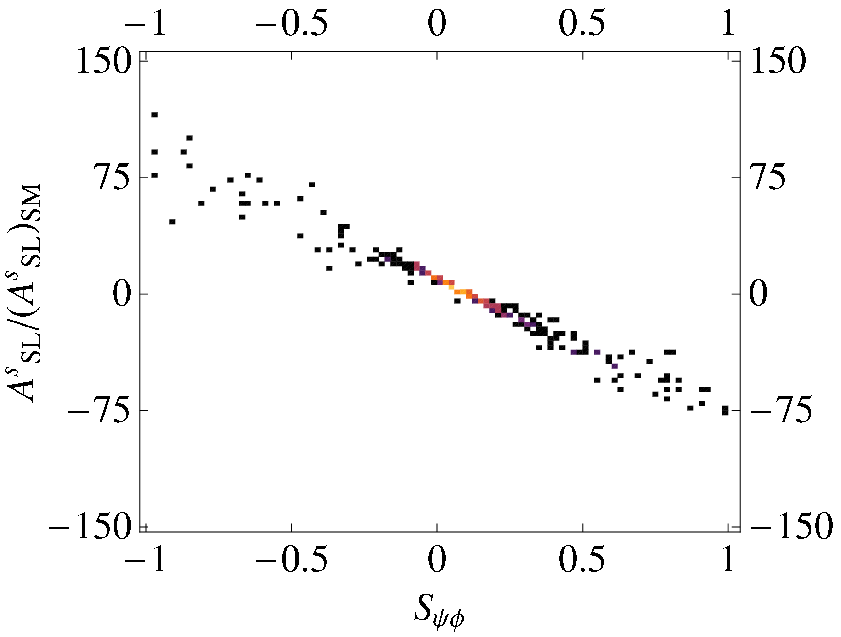,scale=.75}
\end{minipage}
\begin{minipage}{1cm}
\epsfig{file=bar1000.eps,scale=.6}
\end{minipage}
\end{center}
\caption{\label{fig:ASL-const}\it left: $A^s_\text{SL}$, normalised to its SM value, as a function of $S_{\psi\phi}$. In addition to the requirement of correct quark masses and CKM mixings, also the available $\Delta F=2$ constraints are imposed. right: The same, but in addition the condition $\Delta_\text{BG}(\eps_K)<20$ is imposed.  }
\end{figure}

\begin{figure}
\begin{center}
\begin{minipage}{7cm}
\epsfig{file=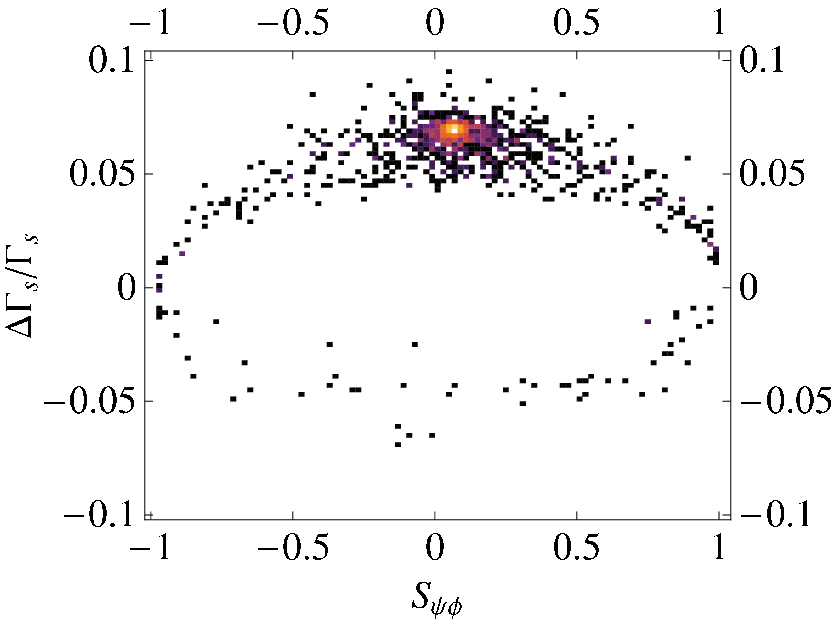,scale=.75}
\end{minipage}
\begin{minipage}{7cm}
\epsfig{file=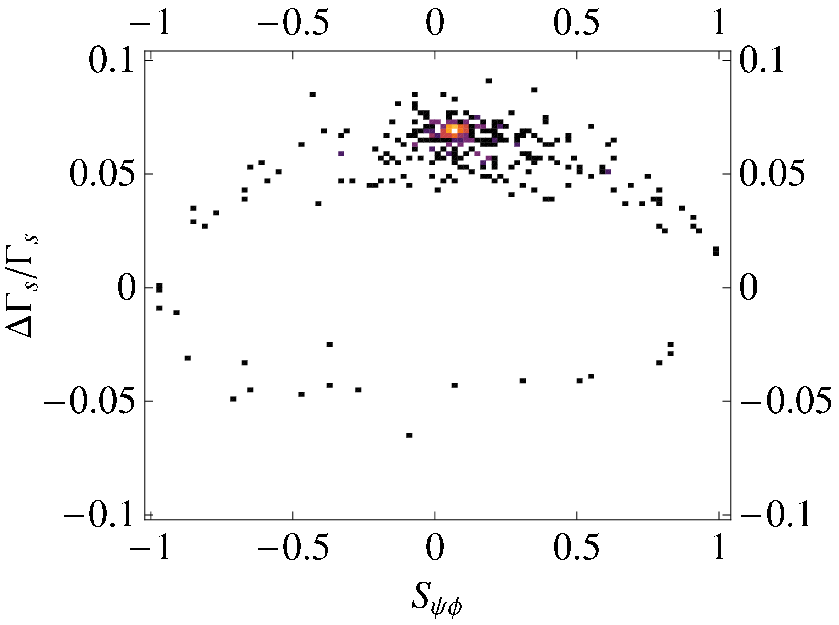,scale=.75}
\end{minipage}
\begin{minipage}{1cm}
\epsfig{file=bar1000.eps,scale=.6}
\end{minipage}
\end{center}
\caption{\label{fig:DGs-const}\it left: $\Delta\Gamma_s/\Gamma_s$ as a function of $S_{\psi\phi}$. In addition to the requirement of correct quark masses and CKM mixings, also the available $\Delta F=2$ constraints are imposed. right: The same, but in addition the condition $\Delta_\text{BG}(\eps_K)<20$ is imposed. }
\end{figure}

Having convinced ourselves that in principle it is possible to obtain agreement with the available $\Delta F=2$ data, we are ready to perform a {\it simultaneous} analysis of all available constraints.  To this end we now impose all $\Delta F=2$ constraints on the RS parameter space, as described in Step 3. The points we show in the subsequent Figures \ref{fig:ASL-const} and \ref{fig:DGs-const} are consistent with the $\Delta F=2$ data and thus fully realistic. In order to maintain naturalness of the theory, the plots in the right panels of these figures fulfil the additional constraint $\Delta_\text{BG}(\eps_K)<20$.

In Fig.~\ref{fig:ASL-const} we show the semileptonic CP-asymmetry $A^s_\text{SL}$ as a function of $S_{\psi\phi}$. We observe that while values of these asymmetries close to the SM ones turn out to be most likely, being a consequence of the {generic relation $|(M^s_{12})_\text{KK}| \sim |(M^s_{12})_\text{SM}|$ observed in Fig.~\ref{fig:MKs12}}, we find that the full range of new physics phases $\varphi_{B_s}$ is possible, so that $-1 <  S_{\psi\phi} < 1$ compared to the SM value $(S_{\psi\phi})_\text{SM}\sim 0.04$, and also $A^s_\text{SL}$ can be enhanced by more than two orders of magnitude relative to its SM value. In addition we observe that the model-independent correlation pointed out in~\cite{Ligeti:2006pm} and verified explicitly in the LHT model in \cite{Blanke:2006sb} turns out to be valid as well in the RS model in question. Comparing the left and right panel with each other we find that the imposition of the naturalness constraint  $\Delta_\text{BG}(\eps_K)<20$ does not qualitatively modify the results obtained, although the overall number of parameter points shown in the plots of course decreases. 

Finally in Fig.~\ref{fig:DGs-const} we show  the width difference $\Delta\Gamma_s/\Gamma_s$ as a function of $S_{\psi\phi}$. We observe that due to the correlation between these two observables, a future more accurate measurement of $\Delta\Gamma_s/\Gamma_s$ {could} help to exclude large values of $S_{\psi\phi}$. 
Again, comparing the left and right panel with each other we find that the overall number of parameter points shown in the plots decreases when imposing $\Delta_\text{BG}(\eps_K)<20$, but the result is not 
 qualitatively modified.

{
\boldmath
\subsubsection{Generic Bound on $M_\text{KK}$}
\unboldmath

So far in our numerical analysis we have fixed the scale $f$ to $1\tev$, corresponding to the KK {gauge boson} mass $M_\text{KK}\simeq
2.45\tev$, as we were mainly interested in studying the effects on $\Delta F=2$ observables of KK {modes} that lie in the reach of LHC. We have found that while it is possible to fulfil all existing constraints, in particular the one from $\eps_K$ even without significant fine-tuning of parameters, we observed that generically a significant amount of fine-tuning is required in order to keep $\eps_K$ in agreement with experiment.

Finally, motivated by the findings in \cite{Csaki:2008zd}, we aim to derive a generic lower bound  on the KK scale $M_\text{KK}$. In order to achieve this we impose the constraint that the \emph{average} fine-tuning required to obtain acceptable values for $\eps_K$ should not exceed a certain value, i.\,e. $\Delta_\text{BG} (\eps_K)_\text{av.} < 10$ or 20. 

In addition to our previously performed scan over the 5D Yukawa couplings and bulk mass parameters, we now take also $f$, or equivalently $M_\text{KK}$, as a free parameter. Fig.~\ref{fig:KKbound} shows the average required fine-tuning in $\eps_K$, {obtained by taking the arithmetic mean of $\Delta_\text{BG}(\eps_K)$ of those points that fulfil the $\eps_K$ constraint within $\pm 30\%$,} as a function of $M_\text{KK}$. We observe that  $\Delta_\text{BG}(\eps_K)$ decreases roughly as $ 1/M_\text{KK}^2$, as expected from the dependence of the KK {gauge boson contributions} to $M_{12}$. {As we have seen in Fig.~\ref{fig:epsK-tuning} the average fine-tuning for points that lie in the generic region for $\eps_K$ is around 20. Therefore, imposing then as naturalness constraint
$\Delta_\text{BG} (\eps_K)_\text{av.} < 20$ we obtain as lower {bound} on the KK scale 
\be
M_\text{KK}\gsim 18\tev\,,
\ee
in rough accordance with the result of \cite{Csaki:2008zd}. We note that the bound in the latter paper has been obtained by the requirement {that the average value for the Wilson coefficients respect the model independent
bounds. This gives the same result as requiring the generic prediction to be within the bounds.}

\begin{figure}
\begin{center}
\epsfig{file=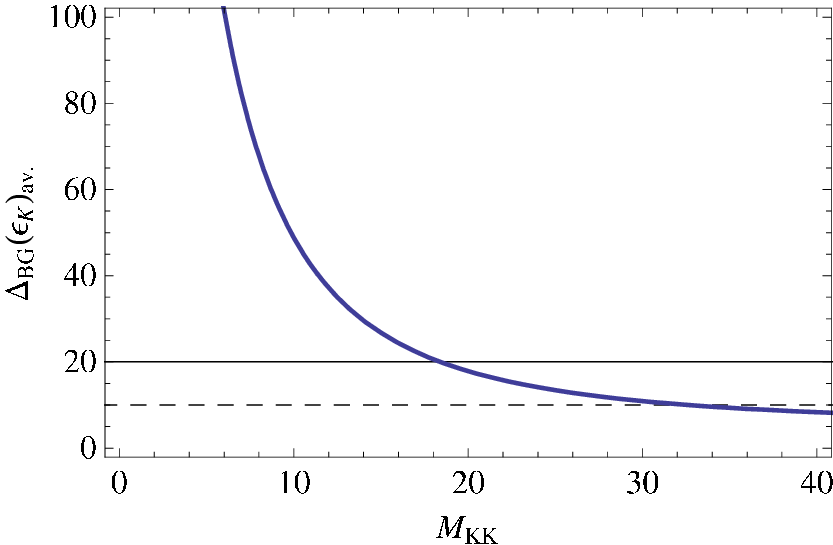}
\end{center}
\caption{\label{fig:KKbound}\it The average required fine-tuning in $\eps_K$ as a function of the KK scale $M_\text{KK}$.}
\end{figure}

If we were to impose instead the more stringent constraint $\Delta_\text{BG} (\eps_K)_\text{av.} < 10$, that is often adopted in the literature, we would find the even stronger constraint
\be
M_\text{KK}\gsim 30\tev\,.
\ee}

Still we would like to stress again, that although this bound can be considered as a naturalness constraint on the theory coming from $\eps_K$, we have found regions of parameter space which yield $\eps_K$ in rough agreement with experiment without any significant fine-tuning for a KK scale as low as $2.5\tev$. {Note however that sub-leading {contributions} like the radiatively induced brane kinetic terms for the fermions are expected to dominate in case of accidental numerical cancellations of the leading terms. 

Thus a natural solution to the ``$\eps_K$ problem'' with KK {gauge bosons} in the reach of the LHC cannot be excluded although we expect it to be radiatively unstable.

}

\newsection{Conclusions}\label{sec:conc}

In the present paper we have performed for the first time the full renormalisation group analysis at the NLO level of the most interesting $\Delta F=2$ observables within the $SU(3)_c\times SU(2)_L\times SU(2)_R\times U(1)_X\times P_{LR}$ model, {including both KK gluon and EW gauge boson contributions}. The protective custodial and $P_{LR}$ symmetries in such models {allow for consistency with EW precision tests for KK scales} as low as $M_\text{KK}\simeq (2-3)\tev$ that are in the reach of {the} LHC. As pointed out in \cite{Csaki:2008zd} for an anarchic structure of the 5D Yukawa couplings much higher KK scales in the ballpark of $(10-20)\tev$ are required in order to satisfy the $\varepsilon_K$-constraint in the presence of KK gluon exchanges. Our detailed analysis confirms these finding, but having at hand more accurate formulae  allows for a quantitative estimate of the fine-tuning in the 5D Yukawa couplings required to reproduce the quark masses and CKM parameters and simultaneously {obtain consistency} of the model with $\varepsilon_K$ and other $\Delta F=2$ observables for $M_\text{KK}\simeq (2-3)\tev$.

The main messages from our analysis are as follows:

\begin{enumerate}
\item While generally $\varepsilon_K$ values turn out to be significantly larger than its experimental value, we {find} regions in parameter space in which the experimental value of $\varepsilon_K$ can be reproduced without large fine-tuning. {The situation is different for the other $\Delta F=2$ observables, where the experimental constraints are naturally fulfilled without significant fine-tuning.}\label{1}

\item {Very interestingly the EW tree level contributions to $\Delta F=2$ observables mediated by new $Z_H$ and $Z'$ weak gauge bosons, while subleading in the case of $\eps_K$ and $\Delta M_K$, turn out to be of  roughly the same size as the KK gluon contributions in the case of $B_{d,s}$ physics observables.
The $Z$ contributions are of $\ord(v^4/M_\text{KK}^4)$ and moreover further suppressed by the custodial protection of {$Zd^i_L\bar d^j_L$}.}\label{5}

\item The amount of fine tuning required to satisfy the $\Delta F=2$ constraints in  $B_d^0-\bar B_d^0$ and  $B_s^0-\bar B_s^0$ systems is considerably smaller than in the case of $\Delta M_K$ and $\varepsilon_K$. This is partly due to the fact that the role of the dangerous $\mathcal{Q}_{LR}$ operators in $\Delta B=2$ transitions turns out to be significantly less important than in $\Delta S=2$ transitions, so that the contributions of the operators $\mathcal{Q}_{LL}$ and $\mathcal{Q}_{LR}$ to the $\Delta B=2$ observables are of the same order.  \label{2}

\item The contributions of KK {gauge boson} tree level exchanges involving new flavour and CP-violating interactions allow not only to satisfy all existing $\Delta F=2$ constraints but also to remove a number of tensions experienced lately by the SM, observed in particular in $\varepsilon_K$, $S_{\psi K_S}$ and $S_{\psi \phi}$ \cite{Buras:2008nn,Lunghi:2008aa,Bona:2008jn,Lenz:2006hd}.\label{3}

\item Most interestingly the model allows naturally for $S_{\psi \phi}$ as high as 0.4 that is hinted at by the most recent CDF and  D{\O} data \cite{Aaltonen:2007he,:2008fj,Brooijmans:2008nt} and by an order of magnitude larger than the SM expectation, $S_{\psi \phi}\simeq 0.04$. The strong correlation between $S_{\psi \phi}$ and $A^s_\text{SL}$ shown in Fig.~\ref{fig:ASL-const} implies then a spectacular departure of the latter observable from its tiny SM value. \label{4}

\item {The effects of the mixing} of the heavy KK quarks with the SM quarks turns out not to be very important in particular in view of  many parameters present in the model. As the fermion representations in the model in question are rather complicated \cite{Blanke:2008aa} and these fermions do not contribute at tree level to $\Delta F=2$ processes we leave a detailed analysis of these effects to a separate publication. \label{6}

\item As a by-product we analysed the connection of RS models to the Froggatt-Nielsen scenario and provided analytic formulae for the effective flavour mixing matrices in terms of the fundamental 5D parameters. We also presented a new useful parameterisation of the 5D Yukawa coupling matrices, taking into account only physical parameters.

\end{enumerate}

Our detailed analysis of {rare $K$ and $B$ decays in the model} in question
is presented in \cite{BBDGG}.

\subsection*{Acknowledgements}

We thank Michaela Albrecht and Katrin Gemmler for very useful discussions and participation in the early stages of this work, and {Csaba Csaki,} Uli Haisch, Tillmann Heidsieck and Diego Guadagnoli for useful discussions.
This research was partially supported by {the Graduiertenkolleg GRK 1054, the Deutsche Forschungsgemeinschaft (DFG) under contract BU 706/2-1}, {the DFG Cluster of Excellence `Origin and Structure of the Universe' and by} the German Bundesministerium f{\"u}r Bildung und
Forschung under contract 05HT6WOA.

\appendix

\newsection{Additional Details on Quark Masses and Flavour Mixing Matrices}\label{app:A}

Explicit expressions for  $\omega^d_{ij}$ and $\rho^d_{ij}$ in \eqref{eq:DLij}--\eqref{eq:alphaij} are given as follows:
\begin{gather}
\omega^d_{ii} =1\,,\qquad 
\omega^d_{12} = \frac{\lambda^d_{33}\lambda^d_{12}-\lambda^d_{13}\lambda^d_{32}}{\lambda^d_{22}\lambda^d_{33}-\lambda^d_{23}\lambda^d_{32}}\,,\qquad
\omega^d_{13} =  \frac{\lambda^d_{13}}{\lambda^d_{33}}\,,     \qquad
\omega^d_{23} =  \frac{\lambda^d_{23}}{\lambda^d_{33}}\,,      \\
\omega^d_{21} =  -\left(\omega^d_{12} \right)^* \,,    \qquad
\omega^d_{31} =   -\left(\omega^d_{13} \right)^* -\left(\omega^d_{23} \right)^* \omega^d_{21} \,,      \qquad
\omega^d_{32} = -\left(\omega^d_{23} \right)^* \,. 
\end{gather}
\begin{gather}
\rho^d_{ii} = 1\,,\qquad
\rho^d_{12} = \left( \frac{\lambda^d_{33}\lambda^d_{21}-\lambda^d_{31}\lambda^d_{23}}{\lambda^d_{22}\lambda^d_{33}-\lambda^d_{23}\lambda^d_{32}} \right)^*\,,\qquad
\rho^d_{13} = \left( \frac{\lambda^d_{31}}{\lambda^d_{33}}\right)^*\,,     \qquad
\rho^d_{23} = \left( \frac{\lambda^d_{32}}{\lambda^d_{33}}\right)^*\,,      \\
\rho^d_{21} =  -\left(\rho^d_{12} \right)^* \,,    \qquad
\rho^d_{31} =   -\left(\rho^d_{13} \right)^* -\left(\rho^d_{23} \right)^* \rho^d_{21} \,,      \qquad
\rho^d_{32} = -\left(\rho^d_{23} \right)^* \,. 
\end{gather}
The expressions for $\omega^u_{ij}$ and $\rho^u_{ij}$, that enter the formulae for $\mathcal{U}_{L,R}$, are obtained by replacing ``$d$'' by ``$u$''.

\newsection{Details on Electroweak Contributions}\label{app:B}

In the case of $A^{(1)}$, $\varepsilon_{L,R}(i)$ in (\ref{varepsilon}) is replaced by $(i=1,2,3)$
\be
\varepsilon_{L,R}(i)(A^{(1)}) = {Q_\text{em}} e^\text{4D}\frac{1}{L}\int_0^L dy\,e^{ky} \left[f^{(0)}_{L,R}(y,c_\Psi^i)\right]^2 g(y)\,,
\label{eq:B.1}
\ee
with $g(y)$ being the gauge KK shape function of $A^{(1)}$. Then the $3\times 3$ matrices  $\hat\Delta_{L,R}(A^{(1)})$ are defined by
\bea
\hat\Delta_{L,R}(A^{(1)})=\mathcal{D}_{L,R}^\dagger\hat\varepsilon_{L,R}(A^{(1)})\mathcal{D}_{L,R}\,,
\label{eq:B.2}
\eea
with $\hat\varepsilon_{L,R}(A^{(1)})$ diagonal matrices analogous to (\ref{eq:3.10}) with the diagonal elements given by $\varepsilon_{L,R}(i)(A^{(1)})$ in (\ref{eq:B.1}). Formula (\ref{eq:B.2}) allows then to find $\Delta_{L,R}^{sd}(A^{(1)})$, $\Delta_{L,R}^{bd}(A^{(1)})$ and $\Delta_{L,R}^{bs}(A^{(1)})$.

In order to give the expressions for $\hat\Delta_{L,R}(Z^{(1)})$ and $\hat\Delta_{L,R}(Z_X^{(1)})$
we introduce
\bea\label{eq:B.5}
\varepsilon_L(i)(Z^{(1)})&=&g_{Z,L}^\text{4D}
\frac{1}{L}\int_0^L dy\,e^{ky} \left[f^{(0)}_{L}(y,c_\Psi^i)\right]^2 g(y)
\\
\varepsilon_R(i)(Z^{(1)})&=&g_{Z,R}^\text{4D}
\frac{1}{L}\int_0^L dy\,e^{ky} \left[f^{(0)}_{R}(y,c_\Psi^i)\right]^2 g(y)
\\
\varepsilon_L(i)(Z_X^{(1)})&=&\kappa_1^\text{4D}
\frac{1}{L}\int_0^L dy\,e^{ky} \left[f^{(0)}_{L}(y,c_\Psi^i)\right]^2\tilde g(y)
\\
\varepsilon_R(i)(Z_X^{(1)})&=&\kappa_5^\text{4D}
\frac{1}{L}\int_0^L dy\,e^{ky} \left[f^{(0)}_{R}(y,c_\Psi^i)\right]^2 \tilde g(y)
\label{eq:B.8}
\eea
with $\tilde g(y)$ being the shape function of $Z_X^{(1)}$, that differs from $g(y)$ due to the different boundary condition on the UV brane.
Further
\bea
g_{Z,L}^\text{4D}&=&\frac{g^\text{4D}}{\cos\psi}\left(-\frac{1}{2}+\frac{1}{3}\sin^2\psi\right)\,,\\
g_{Z,R}^\text{4D}&=&\frac{g^\text{4D}}{\cos\psi}\left(\frac{1}{3}\sin^2\psi\right)\,,\\
\kappa_1^\text{4D}&=&\frac{g^\text{4D}}{\cos\phi}\left(-\frac{1}{2}-\frac{1}{6}\sin^2\phi\right)\,,\\
\kappa_5^\text{4D}&=&\frac{g^\text{4D}}{\cos\phi}\left(-1+\frac{1}{3}\sin^2\phi\right)
\,.
\eea
 Here $g^\text{4D}$ and $g_X^\text{4D}$ are the $SU(2)_L$ and $U(1)_X$ gauge couplings, respectively. Moreover $\sin^2\psi\approx\sin^2\theta_W$ and $\sin\phi$, $\cos\phi$ as functions of $\psi$ are given by the formulae
\be
\cos\psi=\frac{1}{\sqrt{1+\sin^2\phi}}\,,\qquad
\sin\psi=\frac{\sin\phi}{\sqrt{1+\sin^2\phi}}
\ee
 and can also be found in \cite{Blanke:2008aa}. {Note that {in the gauge KK sector}
\be
g_{Z,L}^\text{4D}-\cos\psi\cos\phi\, \kappa_1^\text{4D} =0
\ee
is at the basis of the protection mechanism for flavour diagonal $Zb_L\bar b_L$ and non-diagonal left-handed down quark couplings to $Z$. See Section \ref{sec:Z} for details.

$\hat\Delta_{L,R}(Z^{(1)})$ and $\hat\Delta_{L,R}(Z_X^{(1)})$ are then defined in  analogy to (\ref{eq:B.2}) through
\be
\hat\Delta_{L,R}(Z^{(1)})=\mathcal{D}_{L,R}^\dagger\hat\varepsilon_{L,R}(Z^{(1)})\mathcal{D}_{L,R}
\ee
 with a similar expression for $Z_X^{(1)}$. $\hat\varepsilon_{L,R}$ are diagonal matrices with their elements given by (\ref{eq:B.5})--(\ref{eq:B.8}).

{
\newsection{Tree Level Flavour Changing Higgs Couplings}\label{app:Higgs}

  In this appendix we estimate the size of the relevant {Higgs} vertices by making use of the mass insertion approximation describing the mixing of fermion zero modes with their heavy KK partners. See also \cite{Casagrande:2008hr} for an alternative derivation.

We start by considering diagrams with {one heavy-light transition on a fermion line (denoted by {\color{black}$+$}).} 
\begin{center}
\begin{picture}(200,80)(0,0)
\ArrowLine(40,40)(0,80)
\Line(40,40)(0,0)
\DashLine(40,40)(80,40){4}
\Vertex(40,40){1.3}
\Text(20,15)[cb]{{\Black{$+$}}}
\Text(23,65)[cb]{{\Black{\small${q^d_L}^j$}}}
\Text(16,-2)[cb]{{\Black{\small${q^d_L}^i$}}}
\Text(42,20.5)[cb]{{\Black{\small$d_R^{(1)k}$}}}
\Text(65,43)[cb]{{\Black{\small$h$}}}
\Text(27,30)[cb]{{\Black{\small$\slash{p\,}$}}}
\Text(165,30)[cb]{{\Black{$\sim \lambda^d_{jk} (\lambda^d)^\dagger_{ki} {\,\frac{e^{kL}}{kL}}f^Q_if^Q_j \frac{v \,m^d_i}{M_\text{KK}^2}$ .}}}
\end{picture}
\end{center}
{As the Higgs vertex in that case contains a $P_R$ projector, while the heavy-light mass insertion comes along with a $P_L$, the leading contribution from the $1/M_\text{KK}$ part of the fermion propagator vanishes, and only the non-leading $ \slash{p\,}/M_\text{KK}^2$ contribution survives. When acting on the external fermion, the additional $\slash{p\,}/M_\text{KK}$ results in the strong chiral suppression $m^d_i/M_\text{KK}$.}

{Let us next consider the case of a heavy-heavy transition in addition to the {heavy-light transition} already considered. Na\"ively one may expect that, as now the external fermions are of different {chirality}, the suppression factor $m^d_i/M_\text{KK}$ is absent, so that such diagrams yield the dominant contribution. However, one finds
\begin{center}
\begin{picture}(200,85)(0,0)
\ArrowLine(40,40)(-5,85)
\Line(40,40)(-5,-5)
\DashLine(40,40)(80,40){4}
\Vertex(40,40){1.3}
\Text(15,17.5)[cb]{{\Black{\small$\slash{p\,}$}}}
\Text(30,32.5)[cb]{{\Black{\small$\slash{p\,}$}}}
\Text(25,21)[cb]{{\Black{$+$}}} 
\Text(30,5)[cb]{{\Black{\small${q^d_L}^{(1)l}$}}}
\Text(10.5,6)[cb]{{\Black{$+$}}} 
\Text(45,22)[cb]{{\Black{\small$d_R^{(1)k}$}}}
\Text(11,-8)[cb]{{\Black{\small$d_R^{i}$}}}
\Text(65,43)[cb]{{\Black{\small$h$}}}
\Text(23,65)[cb]{{\Black{\small${q^d_L}^j$}}}
\Text(175,30)[cb]{{\Black{$\sim  \lambda^d_{jk} (\lambda^d)^\dagger_{kl}(\lambda^d)^\dagger_{li}
{\,\frac{e^{kL}}{kL}} f^d_i f^Q_j \frac{v^2 {m^d_i}^2}{M_\text{KK}^4}$ ,}}}
\end{picture}
\end{center}
i.\,e. this type of contribution is highly suppressed not only  by the $v^2/M_\text{KK}^2$ factor coming from the two mass insertions, but in addition receives a double chiral suppression factor ${m^d_i}^2/M_\text{KK}^2$. The origin of this strong chiral suppression is in fact easy to see: The Higgs boson, being confined to the IR brane, can couple only to that chirality of a given fermion KK mode that obeys a Neumann BC on that brane; this chirality is necessarily the one of the corresponding zero mode. {This implies that again only the $\slash{p}$ dependent parts of the KK fermion propagators contribute.} Evaluating then the Dirac structure of the above diagram one ends up with the result stated above. We note that in the case of a bulk Higgs boson the heavy-heavy transition would contain both fermion chiralities, so that the ${m^d_i}^2/M_\text{KK}^2$ suppression would be absent in that case and this kind of diagram would in fact yield the dominant contribution to flavour changing Higgs couplings.
{Consideration of diagrams with KK fermions contributing simultaneously
 on both external lines does not change this conclusion.
}

We conclude that in the present brane-Higgs scenario Higgs contributions to FCNC processes are negligible in the model in question, which we have also verified numerically.}

\providecommand{\href}[2]{#2}\begingroup\raggedright
\endgroup

\end{document}